\documentclass[prd,aps,floatfix,nofootinbib, 10 pt]{revtex4}

\usepackage{amsmath,graphicx,color,epsfig}

\begin{document}

\title{ Exclusive semileptonic decays of $\Lambda_b \to \Lambda l^{+} l^{-}$
in supersymmetric theories}
\author{M. Jamil Aslam$^{1,2}$}
\author{Yu-Ming Wang$^{1}$}
\author{Cai-Dian L\"{u} $^{1}$}
\affiliation{$^{1}$Institute of High Energy Physics, P.O. Box
918(4), Beijing 100049, China} \affiliation{$^{2}$National Center
for Physics, Quaid-i-Azam University, Islamabad, Pakistan}

\begin{abstract}
The weak decays of $\Lambda_b \to \Lambda l^{+} l^{-}$ ($l=e, \,\,
\mu$) are investigated in Minimal Supersymmetric Standard Model
(MSSM) and also in Supersymmetric (SUSY) SO(10) Grand Unified
Models. In MSSM the special attention is paid to the Neutral Higgs
Bosons (NHBs) as they make quite a large contribution in exclusive
$B \to X_{s} l^{+} l^{-}$ decays at large $\tan{\beta}$ regions of
parameter space of SUSY models, since part of SUSY contributions is
proportional to $\tan^{3}{\beta}$. The analysis of decay rate,
forward-backward asymmetries, lepton polarization asymmetries and
the polarization asymmetries of $\Lambda $ baryon in $\Lambda_b \to
\Lambda l^{+} l^{-}$ show that the values of these physical
observables are greatly modified by the effects of NHBs. In SUSY
SO(10) GUT model, the new physics contribution comes from the
operators which are induced by the NHBs penguins and also from the
operators having chirality opposite to that of the corresponding SM
operators. SUSY SO(10) effects show up only in the decay $\Lambda_b
\to \Lambda +\tau^{+} \tau^{-}$ where the longitudinal and
transverse lepton polarization asymmetries are deviate significantly
from the SM value while the effects in the decay rate,
forward-backward asymmetries and polarization asymmetries of final
state $\Lambda$ baryon are very mild. The transverse lepton
polarization asymmetry in $\Lambda_b \to \Lambda +\tau^{+} \tau^{-}$
is almost zero in SM and in MSSM model. However, it can reach to
$-0.1$ in SUSY SO(10) GUT model and could be seen at the future
colliders; hence this asymmetry observable will provide us useful
information to probe new physics and discriminate between different
models.
\end{abstract}

\maketitle

%\vspace*{1.0cm}

%\vspace*{1.0cm}

%\date{\today}

%\newpage

\section{Introduction}

From last decade, rare decays induced by flavor changing neutral
currents (FCNCs) $b \to s\, l^{+} l^{-}$ have become the main focus
of the studies due to the CLEO measurement of the radiative decay $b
\to s\gamma$ \cite{CLEO1}. In the standard model (SM) these decays
are forbidden at tree level and can only be induced by
Glashow-Iliopoulos-Maiani mechanism \cite{Glashow:1970gm} via loop
diagrams. Hence, such decays will provide helpful information about
the parameters of Cabbibo-Kobayashi-Maskawa (CKM) matrix \cite{CKM
1, CKM 2} elements as well as various hadronic form factors. In the
literature there have been intensive studies on the exclusive decays
$B \to P(V, A) \, l^{+} l^{-}$ \cite{b to s in theory 16,b to s in
theory 17,b to s in theory 18,b to s in theory 19,b to s in theory
20,b to s in theory 21,jamil1}  both in the SM and beyond, where the
notions $P, V$ and $A$ denote the pseudoscalar, vector and axial
vector mesons respectively.

It is generally believed that supersymmetry (SUSY) is not only one
of the strongest competitor of the SM but is also the most promising
candidate of new physics. The reason is that it offers a unique
scheme to embed the SM in a more fundamental theory where many
theoretical problems such as gauge hierarchy, origin of mass, and
Yukawa couplings can be resolved. One direct way to search for SUSY
is to discover SUSY particles at high energy colliders, but
unfortunately, so far no SUSY particles have been found. Another way
is to search for its effects through indirect methods. The
measurement of invariant mass spectrum, forward-backward asymmetry
and polarization asymmetries are the suitable tools to probe new
physics effects. For most of the SUSY models, the SUSY contributions
to an observable appear at loop level due to the $\ R$-parity
conservation. Therefore, it has been realized for a long time that
rare processes can be used as a good probe for the searches of SUSY,
since in these processes the contributions of SUSY and SM arises at
the same order in perturbation theory \cite{Yan}.

Motivated from the fact that in two Higgs doublet model and in other
SUSY models, Neutral Higgs Bosons (NHBs) could contribute largely to
the inclusive processes $B \to X_{s} l^{+} l^{-}$, as part of
supersymmetric contributions is proportional to the
$\tan^{3}{\beta}$ \cite{Huang}. Subsequently, the physical
observables, like branching ratio and forward-backward asymmetry, in
the large $\mathrm{\tan{\beta}}$ region of parameter space in SUSY
models can be quite different from that in the SM. In addition,
similar effects in exclusive $B \to K (K^{*})\,\ l^{+} l^{-}$ decay
modes are also investigated \cite{Yan}, where the analysis of decay
rates, forward-backward asymmetries and polarization asymmetries of
final state lepton indicates the significant role of NHBs. It is
believed that physics beyond the SM is essential to explain the
problem of  neutrino oscillation. To this
purpose, a number of SUSY SO(10) models have been proposed in the literature %
\cite{Babu,Mahanthappa,Gut,Senjanovic}. One such model is the SUSY
SO(10) Grand Unified Models (GUT), in which there is a complex
flavor non-diagonal down-type squark mass matrix
element of 2nd and 3rd generations of order one at the GUT scale \cite%
{Gut}. This can induce large flavor off-diagonal coupling such as
the coupling of gluino to the quark and squark which belong to
different generations. These couplings are in general complex and
may contribute to the process of flavor changing neutral currents
(FCNCs). The above analysis of physical observables in $B \to K
(K^{*})\,\ l^{+} l^{-}$ decay is extended in SUSY SO(10) GUT model
in Ref.  \cite{Li}. It is believed that the effects of the
counterparts of usual chromo-magnetic and electromagnetic dipole
moment operators as well as semileptonic operators with opposite
chirality are suppressed by $m_{s}/m_{b}$ in the SM, but in SUSY
SO(10) GUTs their effect can be significant, since
$\delta^{dRR}_{23}$ can be as large as 0.5 \cite{Li,Gut}. Apart from
this, $\delta^{dRR}_{23}$ can induce new operators as the
counterparts of usual scalar operators in SUSY models due to NHB
penguins with gluino-down type squark propagator in the loop. It has
been shown \cite{Li} that the forward-backward asymmetries as well
as the longitudinal and transverse decay widths of $B \to K
(K^{*})\,\ l^{+} l^{-}$ decay, are sensitive to these NHBs effect in
SUSY SO(10) GUT model which can be detected in the future $B$
factories.

Compared to the $B$ meson decays, the investigations of FCNC
$b\rightarrow s$ transition for bottom baryon decays  $\Lambda
_{b}\rightarrow \Lambda l^{+}l^{-}$ are much behind because more
degrees of freedom are involved in the bound state of baryon system
at the quark level. From the experimental point of view, the only
drawback of bottom baryon decays is that the production rate of
$\Lambda _{b}$ baryon in $b$ quark hadronization is about four times
less than that of the $B$ meson. Theoretically, the major interest
in baryonic decays can be attributed to the fact that they can offer
a unique ground to extract the helicity structure of the effective
Hamiltonian for  $b\rightarrow s$ transition in the SM and beyond,
which is lost in the hadronization of mesonic case. The key issue in
the study of exclusive baryonic decays is to properly evaluate the
hadronic matrix elements for $\Lambda _{b} \to \Lambda$, namely the
transition form factors which are obviously governed by
non-perturbative QCD
dynamics. Currently, there has been some studies in the literature on $%
\Lambda _{b}\rightarrow \Lambda $ transition form factors in
different models including pole model (PM) \cite{Mannel}, covariant
oscillator quark model (COQM)\cite{Mohanta}, MIT bag model
(BM)\cite{Cheng} and
non-relativistic quark model \cite{Cheng 2}, QCD sum rule approach (QCDSR) %
\cite{Huang1}, perturbative QCD (pQCD) approach \cite{HLLW} and also
in the light-cone sum rules approach (LCSR) \cite{YuMing}. Using
these form factors, the physical observables like decay rates,
forward-backward asymmetries and polarization asymmetries of
$\Lambda $ baryon as well as of the final state leptons in $\Lambda
_{b}\rightarrow \Lambda l^{+}l^{-}$ were studied in great details in
the literature \cite{c.q. geng 1,c.q. geng 2,c.q. geng 3,c.q. geng
4, Aliev 1,Aliev 2,Aliev 3,Aliev 4}. It is pointed out that these
observables are very sensitive to the new physics, for instance, the
polarization asymmetries of $\Lambda $ baryon in $\Lambda
_{b}\rightarrow \Lambda l^{+}l^{-} $ decays heavily depend on the
right handed current, which is much suppressed in the SM \cite{c.q.
geng 4}.

In this paper, we will investigate the exclusive decay $\Lambda
_{b}\rightarrow \Lambda l^{+}l^{-}$ ( $l$ = $\mu $, $\tau $) both in
the Minimal Supersymmetric Standard Model (MSSM) as well as in the
SUSY SO(10) GUT model \cite{Gut}. We evaluate the branching ratios,
forward-backward asymmetries, lepton polarization asymmetries and
polarization asymmetries of $\Lambda $ baryon with special emphasis
on the effects of NHBs in MSSM. It is pointed out that different
source of the vector current could manifest themselves in different
regions of phase space. For low value of momentum transfer, the
photonic penguin dominates, while the $Z$
penguin and $W$ box become important towards high value of momentum transfer %
\cite{Yan}. In order to search the region of momentum transfer with large
contributions from NHBs, the above decay in certain large $%
\mathrm{{tan{\ \beta}}}$ region of parameter space has been analyzed
in SuperGravity (SUGRA) and M-theory inspired models \cite{CSHuang}.
We extend this analysis to the SUSY SO(10) GUT model \cite{Yan},
where there are some primed counterparts of the usual SM operators.
For instance, the counterparts of usual operators in $\ B\rightarrow
\ X_{s}\ \gamma $ decay are suppressed by $m_{s}/m_{b}$ and
consequently negligible in the SM because they have opposite
chiralities. These operators are also suppressed in Minimal Flavor
Violating (MFV) models \cite{Bobeth,Wu}, however, in SUSY SO(10) GUT
model their effects can be significant. The reason is that the
flavor non-diagonal squark mass matrix elements are the free
parameters and some of them have
significant effects in rare decays of $B$ mesons %
\cite{Lunghi}. In our numerical analysis for $\Lambda
_{b}\rightarrow \Lambda l^{+}l^{-}$ decays, we shall use the results
of the form factors calculated by LCSR approach in Ref.
\cite{YuMing}, and the values of the relevant Wilson coefficient for
MSSM and SUSY SO(10) GUT models are borrowed from Ref. \cite{Yan,
Li}. The effects of SUSY contributions to the decay rate and zero
position of forward-backward asymmetry are also explored in this
work. Our results show that not only the decay rates are sensitive
to the NHBs contribution but the zero point of the forward-backward
asymmetry also shifts remarkably. It is known that the hadronic
uncertainties associated with the form factors and other input
parameters have negligible effects on the lepton polarization
asymmetries and polarization asymmetries of $\Lambda $ baryon in
$\Lambda _{b}\rightarrow \Lambda l^{+}l^{-}$ decays. We have also
studied these asymmetries in the SUSY models mentioned above and
found that the effects of NHBs are quite significant in some regions
of parameter space of SUSY.

The paper is organized as follows. In Sec. II, we present the
effective Hamiltonian for the dilepton decay $\Lambda_b \to \Lambda
l^{+} l^{-}$. Section III contains the definitions and numbers of
the form factors for the said decay using the LCSR approach. In Sec.
IV we present the basic formulas of physical observables like decays
rate, forward-backward asymmetries (FBAs) and polarization
asymmetries of lepton and that of the $\Lambda$ baryon in $\Lambda_b
\to \Lambda l^{+} l^{-}$. Section V is devoted to the numerical
analysis of these observables and the brief summary and concluding
remarks are given in Sec. VI.

\section{Effective Hamiltonian}

After integrating out the heavy degrees of freedom in the full theory, the
general effective Hamiltonian for $\ b\rightarrow \ sl^{+}l^{-}$ in SUSY
SO(10) GUT model, can be written as \cite{Li}
\begin{eqnarray}
H_{eff} &=&-\frac{G_{F}}{2\sqrt{2}}V_{tb}V_{ts}^{\ast }\bigg[{%
\sum\limits_{i=1}^{2}}C_{i}({\mu })O_{i}({\mu })+{\sum\limits_{i=3}^{10}(}%
C_{i}({\mu })O_{i}({\mu })+C_{i}^{\prime }({\mu })O_{i}^{\prime }({\mu }))
\notag \\
&&+\sum\limits_{i=1}^{8}{(}C_{Q_i}({\mu })Q_{i}({\mu })+C_{Q_i}^{\prime }({%
\mu })Q_{i}^{\prime }({\mu }))\bigg],  \label{effective haniltonian 1}
\end{eqnarray}%
where $O_{i}({\mu })$ $(i=1,\ldots ,10)$ are the four-quark operators and $%
C_{i}({\mu })$ are the corresponding Wilson\ coefficients at the energy
scale ${\mu }$ \cite{Goto}. Using renormalization group equations to resume
the QCD corrections, Wilson coefficients are evaluated at the energy scale ${%
\mu =m}_{b}$. The theoretical uncertainties associated with the
renormalization scale can be substantially reduced when the
next-to-leading-logarithm corrections are included \cite{Buchalla}.
The new operators $Q_{i}({\mu })$ $(i=1,\ldots ,8)$ come   from the
NHBs exchange diagrams, whose manifest forms and corresponding
Wilson coefficients can be found in \cite{Ewerth,Feng}. The primed
operators are the counterparts of the unprimed operators, which can
be obtained by flipping the chiralities in the corresponding
unprimed operators. It needs to point out that these primed
operators will appear only in SUSY SO(10) GUT model and are absent
in SM and MSSM \cite{Yan}.

The explicit expressions of the operators responsible for $\Lambda
_{b}\rightarrow \Lambda l^{+}l^{-}$ transition are given by
\begin{eqnarray}
O_{7} &=&\frac{e^{2}}{16\pi ^{2}}m_{b}\left( \bar{s}\sigma _{\mu \nu
}P_{R}b\right) F^{\mu \nu },\,\qquad O_{7}^{\prime }=\frac{e^{2}}{16\pi ^{2}}%
m_{b}\left( \bar{s}\sigma _{\mu \nu }P_{L}b\right) F^{\mu \nu }  \notag \\
O_{9} &=&\frac{e^{2}}{16\pi ^{2}}(\bar{s}\gamma _{\mu }P_{L}b)(\bar{l}\gamma
^{\mu }l),\,\qquad \ \ \ O_{9}^{\prime }=\frac{e^{2}}{16\pi ^{2}}(\bar{s}%
\gamma _{\mu }P_{R}b)(\bar{l}\gamma ^{\mu }l)  \notag \\
O_{10} &=&\frac{e^{2}}{16\pi ^{2}}(\bar{s}\gamma _{\mu }P_{L}b)(\bar{l}%
\gamma ^{\mu }\gamma _{5}l),\,\ \ \ \ \ \ O_{10}^{\prime }=\frac{e^{2}}{%
16\pi ^{2}}(\bar{s}\gamma _{\mu }P_{R}b)(\bar{l}\gamma ^{\mu }\gamma _{5}l)
\notag \\
Q_{1} &=&\frac{e^{2}}{16\pi ^{2}}(\bar{s}P_{R}b)(\bar{l}l),\qquad \qquad \ \
\ \ \ Q_{1}^{\prime }=\frac{e^{2}}{16\pi ^{2}}(\bar{s}P_{L}b)(\bar{l}l)
\notag \\
Q_{2} &=&\frac{e^{2}}{16\pi ^{2}}(\bar{s}P_{R}b)(\bar{l}\gamma _{5}l),\qquad
\ \ \ \ \ \ \ \ Q_{2}^{\prime }=\frac{e^{2}}{16\pi ^{2}}(\bar{s}P_{L}b)(\bar{%
l}\gamma _{5}l)  \label{relvent-operators}
\end{eqnarray}%
with $P_{L,R}=\left( 1\pm \gamma _{5}\right) /2$. In terms of the
above
Hamiltonian, the free quark decay amplitude for $b\rightarrow s$ $%
l^{+}l^{-} $ can be derived as \cite{Huang}:

\begin{eqnarray}
\mathcal{M}(b &\rightarrow &sl^{+}l^{-})=-\frac{G_{F}\alpha}{\sqrt{2}\pi }%
V_{tb}V_{ts}^{\ast }\bigg\{C_{9}^{eff}(\bar{s}\gamma _{\mu }P_L b)(\bar{l}%
\gamma ^{\mu }l)+C_{10}(\bar{s}\gamma _{\mu }P_L b)(\bar{l}\gamma
^{\mu }\gamma
_{5}l)  \notag \\
&&-2m_{b}C_{7}^{eff}(\bar{s}i\sigma _{\mu \nu }\frac{q^{\nu }}{s}P_R b)(\bar{l}%
\gamma ^{\mu }l)+C_{Q_{1}}(\bar{s}P_R b)(\bar{l}l)+C_{Q_{2}}(\bar{s}P_R b)(\bar{l}%
\gamma _{5}l)+(C_{i}(m_{b})\longleftrightarrow C_{i}^{\prime }(m_{b}))\bigg\}
\notag \\
&&  \label{quark-amplitude}
\end{eqnarray}%
where $s=q^{2}$ and $q=p_{\Lambda _{b}}-p_{\Lambda }$ is the
momentum transfer. Due to the absence of $Z$ boson in the effective
theory, the operator $O_{10}$ can not be induced by the insertion of
four-quark operators. Therefore, the Wilson coefficient $C_{10}$
does not renormalize under QCD corrections and hence it is
independent on the energy scale. Moreover, the above quark level
decay amplitude can receive additional contributions from the matrix
element of four-quark operators, $\sum_{i=1}^{6}\langle
l^{+}l^{-}s|O_{i}|b\rangle $, which are usually absorbed into the
effective Wilson coefficient $C_{9}^{eff}(\mu )$. To be more
specific, we can decompose $C_9^{eff}(\mu)$ into the following three
parts \cite{b to s in theory 3,b to s in theory 4,b to s in theory
5,b to s in theory 6,b to s in theory 7,b to s in theory 8,b to s in
theory 9}
\begin{equation*}
C_{9}^{eff}(\mu )=C_{9}(\mu )+Y_{SD}(z,s^{\prime })+Y_{LD}(z,s^{\prime }),
\end{equation*}%
where the parameters $z$ and $s^{\prime }$ are defined as $%
z=m_{c}/m_{b},\,\,\,s^{\prime }=q^{2}/m_{b}^{2}$. $Y_{SD}(z,s^{\prime })$
describes the short-distance contributions from four-quark operators far
away from the $c\bar{c}$ resonance regions, which can be calculated reliably
in the perturbative theory. The long-distance contributions $%
Y_{LD}(z,s^{\prime })$ from four-quark operators near the $c\bar{c}$
resonance cannot be calculated from first principles of QCD and are usually
parameterized in the form of a phenomenological Breit-Wigner formula making
use of the vacuum saturation approximation and quark-hadron duality. The
manifest expressions for $Y_{SD}(z,s^{\prime })$ and $Y_{LD}(z,s^{\prime })$
can be written as \cite{YuMing,c.q. geng 1, c.q. geng 2,c.q. geng 3,c.q.
geng 4}
\begin{eqnarray}
Y_{SD}(z,s^{\prime }) &=&h(z,s^{\prime })(3C_{1}(\mu )+C_{2}(\mu
)+3C_{3}(\mu )+C_{4}(\mu )+3C_{5}(\mu )+C_{6}(\mu ))  \notag \\
&&-\frac{1}{2}h(1,s^{\prime })(4C_{3}(\mu )+4C_{4}(\mu )+3C_{5}(\mu
)+C_{6}(\mu ))  \notag \\
&&-\frac{1}{2}h(0,s^{\prime })(C_{3}(\mu )+3C_{4}(\mu ))+{\frac{2}{9}}%
(3C_{3}(\mu )+C_{4}(\mu )+3C_{5}(\mu )+C_{6}(\mu )),
\end{eqnarray}%
\begin{eqnarray}
Y_{LD}(z,s^{\prime }) &=&\frac{3}{\alpha _{em}^{2}}(3C_{1}(\mu )+C_{2}(\mu
)+3C_{3}(\mu )+C_{4}(\mu )+3C_{5}(\mu )+C_{6}(\mu ))  \notag \\
&&\sum_{j=\psi ,\psi ^{\prime }}\omega _{j}(q^{2})k_{j}\frac{\pi \Gamma
(j\rightarrow l^{+}l^{-})M_{j}}{q^{2}-M_{j}^{2}+iM_{j}\Gamma _{j}^{tot}},
\label{LD}
\end{eqnarray}%
with
\begin{eqnarray}
h(z,s^{\prime }) &=&-{\frac{8}{9}}\mathrm{ln}z+{\frac{8}{27}}+{\frac{4}{9}}x-%
{\frac{2}{9}}(2+x)|1-x|^{1/2}\left\{
\begin{array}{l}
\ln \left| \frac{\sqrt{1-x}+1}{\sqrt{1-x}-1}\right| -i\pi \quad \mathrm{for}{%
{\ }x\equiv 4z^{2}/s^{\prime }<1} \\
2\arctan \frac{1}{\sqrt{x-1}}\qquad \mathrm{for}{{\ }x\equiv
4z^{2}/s^{\prime }>1}%
\end{array}%
\right. ,  \notag \\
h(0,s^{\prime }) &=&{\frac{8}{27}}-{\frac{8}{9}}\mathrm{ln}{\frac{m_{b}}{\mu
}}-{\frac{4}{9}}\mathrm{ln}s^{\prime }+{\frac{4}{9}}i\pi \,\,.
\end{eqnarray}

The non-factorizable effects \cite{b to s 1, b to s 2, b to s 3,NF charm
loop} from the charm loop can bring about further corrections to the
radiative $b\rightarrow s\gamma $ transition, which can be absorbed into the
effective Wilson coefficient $C_{7}^{eff}$. Specifically, the Wilson
coefficient $C^{eff}_{7}$ is given by \cite{c.q. geng 4}
\begin{equation*}
C_{7}^{eff}(\mu )=C_{7}(\mu )+C_{b\rightarrow s\gamma }(\mu ),
\end{equation*}%
with
\begin{eqnarray}
C_{b\rightarrow s\gamma }(\mu ) &=&i\alpha _{s}\bigg[{\frac{2}{9}}\eta
^{14/23}(G_{1}(x_{t})-0.1687)-0.03C_{2}(\mu )\bigg], \\
G_{1}(x) &=&{\frac{x(x^{2}-5x-2)}{8(x-1)^{3}}}+{\frac{3x^{2}\mathrm{ln}^{2}x%
}{4(x-1)^{4}}},
\end{eqnarray}%
where $\eta =\alpha _{s}(m_{W})/\alpha _{s}(\mu )$, $%
x_{t}=m_{t}^{2}/m_{W}^{2}$, $C_{b\rightarrow s\gamma }$ is the absorptive
part for the $b\rightarrow sc\bar{c}\rightarrow s\gamma $ rescattering and
we have dropped out the tiny contributions proportional to CKM sector $%
V_{ub}V_{us}^{\ast }$. In addition, $C_{7}^{\prime eff}(\mu )$ and $%
C_{9}^{\prime eff}(\mu )$ can be obtained by replacing the unprimed Wilson
coefficients with the corresponding prime ones in the above formula.

\section{Matrix elements and form factors in Light Cone Sum Rules}

With the free quark decay amplitude available, we can proceed to calculate
the decay amplitudes for $\Lambda_b\to \Lambda \gamma$ and $\Lambda_b \to
\Lambda l^+l^-$ at hadron level, which can be obtained by sandwiching the
free quark amplitudes between the initial and final baryon states.
Consequently, the following four hadronic matrix elements
\begin{eqnarray}
\langle \Lambda(P)|\bar{s}\gamma_{\mu} b|\Lambda_{b}(P+q)\rangle &,& \,\,\,
\langle \Lambda(P)|\bar{s}\gamma_{\mu}\gamma_5 b|\Lambda_{b}(P+q)\rangle ,
\notag \\
\langle \Lambda(P)|\bar{s}\sigma_{\mu \nu} b|\Lambda_{b}(P+q)\rangle &,&
\,\,\, \langle \Lambda(P)|\bar{s}\sigma_{\mu \nu} \gamma_5
b|\Lambda_{b}(P+q)\rangle,  \label{nonvanishing-melements}
\end{eqnarray}
need to be computed. Generally, the above matrix elements can be
parameterized in terms of the form factors as \cite{c.q. geng 4, Aliev
1,Aliev 2,Aliev 3,Aliev 4}:
\begin{eqnarray}
\langle \Lambda (P)|\bar{s}\gamma _{\mu }b|\Lambda _{b}(P+q)\rangle &=&%
\overline{\Lambda }(P)(g_{1}\gamma _{\mu }+g_{2}i\sigma _{\mu \nu }q^{\nu
}+g_{3}q_{\mu })\Lambda _{b}(P+q),\,\,  \label{vectormatrix element} \\
\langle \Lambda (P)|\bar{s}\gamma _{\mu }\gamma _{5}b|\Lambda
_{b}(P+q)\rangle &=&\overline{\Lambda }(P)(G_{1}\gamma _{\mu }+G_{2}i\sigma
_{\mu \nu }q^{\nu }+G_{3}q_{\mu })\gamma _{5}\Lambda _{b}(P+q),\,\,
\label{axial-vectormatrix element} \\
\langle \Lambda (P)|\bar{s}\sigma _{\mu \nu }b|\Lambda _{b}(P+q)\rangle &=&%
\overline{\Lambda }(P)[h_{1}\sigma _{\mu \nu }-ih_{2}(\gamma _{\mu }q_{\nu
}-\gamma _{\nu }q_{\mu })  \notag \\
&&-ih_{3}(\gamma _{\mu }P_{\nu }-\gamma _{\nu }P_{\mu })-ih_{4}(P_{\mu
}q_{\nu }-P_{\nu }q_{\mu })]\Lambda _{b}(P+q),\,\,
\label{tensor matrix element 1} \\
\langle \Lambda (P)|\bar{s}\sigma _{\mu \nu }\gamma _{5}b|\Lambda
_{b}(P+q)\rangle &=&\overline{\Lambda }(P)[H_{1}\sigma _{\mu \nu
}-iH_{2}(\gamma _{\mu }q_{\nu }-\gamma _{\nu }q_{\mu })  \notag \\
&&-iH_{3}(\gamma _{\mu }P_{\nu }-\gamma _{\nu }P_{\mu })-iH_{4}(P_{\mu
}q_{\nu }-P_{\nu }q_{\mu })]\gamma _{5}\Lambda _{b}(P+q),
\label{tensor matrix element 2}
\end{eqnarray}%
where all the form factors $g_{i}$, $G_{i}$, $h_{i}$ and $H_{i}$ are
functions of the square of momentum transfer $q^{2}$. Contracting Eqs. (\ref%
{tensor matrix element 1}-\ref{tensor matrix element 2}) with the four
momentum $q^{\mu }$ on both side and making use of the equations of motion%
\begin{eqnarray}
q^{\mu }(\bar{\psi}_{1}\gamma _{\mu }\psi _{2}) &=&(m_{1}-m_{2})\bar{\psi}%
_{1}\psi _{2}  \label{eq-motion1} \\
q^{\mu }(\bar{\psi}_{1}\gamma _{\mu }\gamma _{5}\psi _{2}) &=&-(m_{1}+m_{2})%
\bar{\psi}_{1}\gamma _{5 }\psi _{2}  \label{eq-motion}
\end{eqnarray}%
we have
\begin{eqnarray}
\langle \Lambda (P)|\bar{s}i\sigma _{\mu \nu }q^{\nu }b|\Lambda
_{b}(P+q)\rangle &=&\overline{\Lambda }(P)(f_{1}\gamma _{\mu }+f_{2}i\sigma
_{\mu \nu }q^{\nu }+f_{3}q_{\mu })\Lambda _{b}(P+q),\,\,
\label{magneticmatrix element1} \\
\langle \Lambda (P)|\bar{s}i\sigma _{\mu \nu }\gamma _{5}q^{\nu }b|\Lambda
_{b}(P+q)\rangle &=&\overline{\Lambda }(P)(F_{1}\gamma _{\mu }+F_{2}i\sigma
_{\mu \nu }q^{\nu }+F_{3}q_{\mu })\gamma _{5}\Lambda _{b}(P+q),
\label{magneticmatrix element2}
\end{eqnarray}%
with
\begin{eqnarray}
f_{1} &=&{\frac{2h_{2}-h_{3}+h_{4}(m_{\Lambda _{b}}+m_{\Lambda })}{2}}q^{2},
\\
f_{2} &=&{\frac{2h_{1}+h_{3}(m_{\Lambda }-m_{\Lambda _{b}})+h_{4}q^{2}}{2}},
\\
f_{3} &=&{\frac{m_{\Lambda }-m_{\Lambda _{b}}}{q^{2}}}f_{1}, \\
F_{1} &=&{\frac{2H_{2}-H_{3}+H_{4}(m_{\Lambda _{b}}-m_{\Lambda })}{2}}q^{2},
\\
F_{2} &=&{\frac{2H_{1}+H_{3}(m_{\Lambda }+m_{\Lambda _{b}})+H_{4}q^{2}}{2}},
\\
F_{3} &=&{\frac{m_{\Lambda }+m_{\Lambda _{b}}}{q^{2}}}F_{1}.
\end{eqnarray}%
Due to the conservation of vector current, the form factors $f_{3}$ and $%
g_{3}$ do not contribute to the decay amplitude of $\Lambda _{b}\rightarrow
\Lambda l^{+}l^{-}$. To incorporate the NHBs effect one need to calculate
the matrix elements involving the scalar $\bar{s}b$ and the pseudoscalar $%
\bar{s}\gamma _{5}b$ currents, which can be parameterized as
\begin{eqnarray}
\langle \Lambda (P)|\bar{s}b|\Lambda _{b}(P+q)\rangle &=&{\frac{1}{%
m_{b}+m_{s}}}\overline{\Lambda }(P)[g_{1}(m_{\Lambda _{b}}-m_{\Lambda
})+g_{3}q^{2}]\Lambda _{b}(P+q),\,\,  \label{scalarmatrix element} \\
\langle \Lambda (P)|\bar{s}\gamma _{5}b|\Lambda _{b}(P+q)\rangle &=&{\frac{1%
}{m_{b}-m_{s}}}\overline{\Lambda }(P)[G_{1}(m_{\Lambda _{b}}+m_{\Lambda
})-G_{3}q^{2}]\gamma _{5}\Lambda _{b}(P+q).
\label{pseudo-scalar matrix element}
\end{eqnarray}

The various form factors $f_{i}$ and $g_{i}$ appearing in the above
equations are not independent in the heavy quark limit and one can express
them in terms of two independent form factors $\xi _{1}$ and $\xi _{2}$ in
HQET defined by \cite{YuMing}
\begin{equation}
\langle \Lambda (P)|\bar{b}\Gamma s|\Lambda _{b}(P+q)\rangle =\overline{%
\Lambda }(P)[\xi _{1}(q^{2})+ {\not v} \xi _{2}(q^{2})]\Gamma \Lambda
_{b}(P+q),  \label{form factors in HQET}
\end{equation}%
with $\Gamma $ being an arbitrary Lorentz structure and $v_{\mu }$ being the
four-velocity of $\Lambda _{b}$ baryon. Comparing Eqs. (\ref{vectormatrix
element}-\ref{axial-vectormatrix element}), (\ref{magneticmatrix element1}- %
\ref{magneticmatrix element2}) and the Eq. (\ref{form factors in HQET}), one
can arrive at \cite{c.q. geng 4, Aliev 1,Aliev 2,Aliev 3,Aliev 4}
\begin{eqnarray}
f_{1} &=&F_{1}={\frac{q^{2}}{m_{\Lambda _{b}}}}\xi _{2}, \\
f_{2} &=&F_{2}=g_{1}=G_{1}=\xi _{1}+{\frac{m_{\Lambda }}{m_{\Lambda _{b}}}}%
\xi _{2},  \label{relatins of from factors in HQET 1} \\
f_{3} &=&{\frac{m_{\Lambda }-m_{\Lambda _{b}}}{m_{\Lambda _{b}}}}\xi _{2}, \\
F_{3} &=&{\frac{m_{\Lambda }+m_{\Lambda _{b}}}{m_{\Lambda _{b}}}}\xi _{2}, \\
g_{2} &=&G_{2}=g_{3}=G_{3}={\frac{\xi _{2}}{m_{\Lambda _{b}}}}.
\label{relatins offrom factors in HQET 2}
\end{eqnarray}

Due to our poor understanding towards non-perturbative QCD dynamics,
one has to rely on some approaches to calculate the form factors answering for $%
\Lambda_b \to \Lambda$ transition. It is suggested that the soft
non-perturbative contribution to the transition form factor can be
calculated quantitatively in the framework of LCSR approach
\cite{LCSR 1, LCSR 2,LCSR 3,LCSR 4,LCSR 5}, which is a fully
relativistic approach and well rooted in quantum field theory, in a
systematic and almost model-independent way. As a marriage of
standard QCDSR
technique \cite{SVZ 1, SVZ 2,SVZ 3} and theory of hard exclusive process %
\cite{hard exclusive process 1,hard exclusive process 2,hard
exclusive process 3,hard exclusive process 4,hard exclusive process
5,hard exclusive process 6,hard exclusive process 7,hard exclusive
process 8}, LCSR cure the problem of QCDSR applying to the large
momentum transfer by performing the operator product expansion (OPE)
in terms of twist of the relevant operators rather than their
dimension \cite{braun talk}. Therefore, the principal discrepancy
between QCDSR and LCSR consists in that non-perturbative vacuum
condensates representing the long-distance quark and gluon
interactions in the short-distance expansion are substituted by the
light cone distribution amplitudes (LCDAs) describing the
distribution of longitudinal momentum carried by the valence quarks
of hadronic bound system in the expansion of transverse-distance
between partons in the infinite momentum frame.

Considering the distribution amplitude up to twist-6, the form factors for $%
\Lambda _{b}\rightarrow \Lambda l^{+}l^{-}$ have been calculated in \cite%
{YuMing} to the accuracy of leading conformal spin, where the pole model was
also employed to extend the results to the whole kinematical region.
Specifically, the dependence of form factors on transfer momentum are
parameterized as
\begin{equation}
\xi _{i}(q^{2})={\frac{\xi _{i}(0)}{1-a_{1}q^{2}/m_{\Lambda
_{b}}^{2}+a_{2}q^{4}/m_{\Lambda _{b}}^{4}}},
\label{pole model of
form factors}
\end{equation}%
where $\xi _{i}$ denotes the form factors $f_{2}$ and $g_{2}$. The numbers
of parameters $\xi_i(0), \,\, a_1, \,\, a_2$ have been collected in Table %
\ref{di-fit}.
\begin{table}[tbh]
\caption{{}Numerical results for the form factors $f_{2}(0)$, $g_{2}(0)$ and
parameters $a_{1}$ and $a_{2}$ involved in the double-pole fit of eq. (\ref%
{pole model of form factors}) for both twist-3 and twist-6 sum rules with $%
M_{B}^{2}\in \lbrack 3.0,6.0]~\mbox{GeV}^{2}$, $s_{0}=39\pm 1~\mbox{GeV}^{2}$%
.}
\label{di-fit}%
\begin{tabular}{ccc}
\hline\hline
parameter & twist-3 & up to twist-6 \\ \hline
$f_{2}(0)$ & $0.14_{-0.01}^{+0.02}$ & $0.15_{-0.02}^{+0.02}$ \\
{$a_{1}$} & $2.91_{-0.07}^{+0.10}$ & $2.94_{-0.06}^{+0.11}$ \\
{$a_{2}$} & $2.26_{-0.08}^{+0.13}$ & $2.31_{-0.10}^{+0.14}$ \\ \hline
$g_{2}(0)(10^{-2}\mathrm{{GeV^{-1}})}$ & $-0.47_{-0.06}^{+0.06}$ & $%
1.3_{-0.4}^{+0.2}$ \\
{$a_{1}$} & $3.40_{-0.05}^{+0.06}$ & $2.91_{-0.09}^{+0.12}$ \\
{$a_{2}$} & $2.98_{-0.08}^{+0.09}$ & $2.24_{-0.13}^{+0.17}$ \\ \hline\hline
\end{tabular}%
\end{table}
To the leading order and leading power, the other form factors can be
related to these two as
\begin{eqnarray}
F_1(q^2)&=&f_1(q^2)=q^2 g_2(q^2)=q^2 G_2(q^2),  \notag \\
F_2(q^2)&=&f_2(q^2)=g_1(q^2)=G_1(q^2),
\end{eqnarray}
where the form factors $F_3(q^2)$ and $G_3(q^2)$ are dropped out here due to
their tiny contributions.

\section{Formula for Observables}

In this section, we proceed to perform the calculations of some
interesting observables in phenomenology including decay rates,
forward-backward asymmetry, polarization asymmetries of final state
lepton and of $\Lambda $ baryon. From Eq. (\ref{quark-amplitude}),
it is straightforward to obtain the decay amplitude for $\Lambda
_{b}\rightarrow \Lambda l^{+}l^{-}$ as
\begin{equation}
\mathcal{M}_{\Lambda _{b}\rightarrow \Lambda
l^{+}l^{-}}=-\frac{G_{F}\alpha }{2\sqrt{2}\pi }V_{tb}V_{ts}^{\ast
}\left[ T_{\mu }^{1}(\bar{l}\gamma ^{\mu }l)+T_{\mu
}^{2}(\bar{l}\gamma ^{\mu }\gamma _{5}l)+T^{3}(\bar{l}l)\right],
\label{lambda-amplitude}
\end{equation}%
where the auxiliary functions $T_{\mu }^{1}$, $T_{\mu }^{2}$ and $T^{3}$ are
given by
\begin{eqnarray}
T_{\mu }^{1} &=&\overline{\Lambda }(P)\bigg[\left\{ \gamma _{\mu }\left(
g_{1}-G_{1}\gamma _{5}\right) +i\sigma _{\mu \nu }q^{\nu }\left(
g_{2}-G_{2}\gamma _{5}\right) \right\} C_{9}^{eff}  \notag \\
&&+\left\{ \gamma _{\mu }\left( g_{1}+G_{1}\gamma _{5}\right) +i\sigma _{\mu
\nu }q^{\nu }\left( g_{2}+G_{2}\gamma _{5}\right) \right\} C_{9}^{\prime eff}
\notag \\
&&-2m_{b}/s\left\{ \gamma _{\mu }\left( f_{1}+F_{1}\gamma _{5}\right)
+i\sigma _{\mu \nu }q^{\nu }\left( f_{2}+F_{2}\gamma _{5}\right) \right\}
C_{7}^{eff}  \notag \\
&&-2m_{b}/s\left\{ \gamma _{\mu }\left( f_{1}-F_{1}\gamma
_{5}\right) +i\sigma _{\mu \nu }q^{\nu }\left( f_{2}-F_{2}\gamma
_{5}\right) \right\} C_{7}^{\prime eff}\bigg]\Lambda _{b}\left(
P+q\right), \label{first-aux-function}
\end{eqnarray}%
\begin{eqnarray}
T_{\mu }^{2} &=&\overline{\Lambda }(P)\bigg[\left\{ \gamma _{\mu }\left(
g_{1}-G_{1}\gamma _{5}\right) +i\sigma _{\mu \nu }q^{\nu }\left(
g_{2}-G_{2}\gamma _{5}\right) +\left( g_{3}-G_{3}\gamma _{5}\right) q_{\mu
}\right\} C_{10}  \notag \\
&&+\left\{ \gamma _{\mu }\left( g_{1}+G_{1}\gamma _{5}\right) +i\sigma _{\mu
\nu }q^{\nu }\left( g_{2}+G_{2}\gamma _{5}\right) +\left( g_{3}+G_{3}\gamma
_{5}\right) q_{\mu }\right\} C_{10}^{\prime }  \notag \\
&&-\frac{q_{\mu }}{2m_{l}\left( m_{b}+m_{s}\right) }\left\{ \left(
g_{1}\left( m_{\Lambda _{b}}-m_{\Lambda }\right) +g_{3}q^{2}+G_{1}\left(
m_{\Lambda _{b}}+m_{\Lambda }\right) -G_{3}q^{2}\right) \right\} C_{Q_{2}}
\notag \\
&&-\frac{q_{\mu }}{2m_{l}\left( m_{b}+m_{s}\right) }\left\{ \left(
g_{1}\left( m_{\Lambda _{b}}-m_{\Lambda }\right) +g_{3}q^{2}-G_{1}\left(
m_{\Lambda _{b}}+m_{\Lambda }\right) +G_{3}q^{2}\right) \right\}
C_{Q_{2}}^{\prime }\bigg]\Lambda _{b}\left( P+q\right),\notag \\
\label{second-aux-function}
\end{eqnarray}%
and
\begin{eqnarray}
T^{3} &=&\overline{\Lambda }(P)\bigg[\left\{ \left( g_{1}\left( m_{\Lambda
_{b}}-m_{\Lambda }\right) +g_{3}q^{2}+G_{1}\left( m_{\Lambda
_{b}}+m_{\Lambda }\right) -G_{3}q^{2}\right) \right\} C_{Q_{1}}  \notag \\
&&+\left\{ \left( g_{1}\left( m_{\Lambda _{b}}-m_{\Lambda }\right)
+g_{3}q^{2}-G_{1}\left( m_{\Lambda _{b}}+m_{\Lambda }\right)
+G_{3}q^{2}\right) \right\} C_{Q_{1}}^{\prime }\bigg]\Lambda
_{b}\left( P+q\right).   \label{third-aux-function}
\end{eqnarray}%
It needs to point out that the terms proportional to $q_{\mu }$ in $T_{\mu
}^{1}$ do not contribute to the decay amplitude with the help of the
equation of motion for lepton fields. Besides, one can also find that the
above results can indeed reproduce that obtained in the SM with $%
C_{i}^{\prime }=0$ and $T^{3}=0$.

\subsection{The differential decay rates of $\Lambda _{b}\rightarrow \Lambda
l^{+}l^{-}$}

The differential decay width of $\Lambda _{b}\rightarrow \Lambda l^{+}l^{-}$
in the rest frame of $\Lambda _{b}$ baryon can be written as \cite{PDG},
\begin{equation}
{\frac{d\Gamma ({\Lambda _{b}\rightarrow \Lambda l^{+}l^{-}})}{ds}}={\frac{1%
}{(2\pi )^{3}}}{\frac{1}{32m_{\Lambda _{b}}^{3}}}\int_{u_{min}}^{u_{max}}|{%
\widetilde{M}}_{\Lambda _{b}\rightarrow \Lambda l^{+}l^{-}}|^{2}du,
\label{differential decay width}
\end{equation}%
where $u=(p_{\Lambda }+p_{l^{-}})^{2}$ and $s=(p_{l^{+}}+p_{l^{-}})^{2}$; $%
p_{\Lambda }$, $p_{l^{+}}$ and $p_{l^{-}}$ are the four-momenta vectors of $%
\Lambda $, $l^{+}$ and $l^{-}$ respectively.
${\widetilde{M}}_{\Lambda _{b}\rightarrow \Lambda l^{+}l^{-}}$
denotes the decay amplitude after performing the integration over
the angle between the $l^{-}$ and $\Lambda $ baryon. The upper and
lower limits of $u$ are given by
\begin{eqnarray}
u_{max} &=&(E_{\Lambda }^{\ast }+E_{l}^{\ast })^{2}-(\sqrt{E_{\Lambda
}^{\ast 2}-m_{\Lambda }^{2}}-\sqrt{E_{l}^{\ast 2}-m_{l}^{2}})^{2},  \notag \\
u_{min} &=&(E_{\Lambda }^{\ast }+E_{l}^{\ast
})^{2}-(\sqrt{E_{\Lambda }^{\ast 2}-m_{\Lambda
}^{2}}+\sqrt{E_{l}^{\ast 2}-m_{l}^{2}})^{2}, \label{defination-u}
\end{eqnarray}%
where $E_{\Lambda }^{\ast }$ and $E_{l}^{\ast }$ are the energies of $%
\Lambda $ and $l^{-}$ in the rest frame of lepton pair
\begin{equation}
E_{\Lambda }^{\ast }={\frac{m_{\Lambda _{b}}^{2}-m_{\Lambda }^{2}-s}{2\sqrt{s%
}}},\hspace{1cm}E_{l}^{\ast }={\frac{\sqrt{s}}{2}}.
\label{defination-energy}
\end{equation}
Putting everything together, we can achieve the decay rates and
invariant mass distributions of $\Lambda _{b}\rightarrow \Lambda
l^{+}l^{-}$ with and without long distance contributions as
\begin{eqnarray}
{\frac{d\Gamma }{ds}} &=&\frac{\alpha ^{2}G_{F}^{2}\left| V_{tb}V_{ts}^{\ast
}\right| ^{2}\left( u_{\max }-u_{\min }\right) }{128m_{\Lambda _{b}}^{3}\pi
^{5}}\times  \notag \\
&&\bigg\{(f_{2}m_{l}(f_{2}-g_{2}((m_{\Lambda }+m_{\Lambda
_{b}}))(C_{Q_{1}}C_{7}^{\ast eff}+C_{Q_{1}}^{\ast }C_{7}^{eff})+\frac{m_{l}}{%
2m_{b}}(f_{2}(m_{\Lambda }+m_{\Lambda _{b}})-g_{2}s)(C_{9}^{\ast
eff}C_{Q_{1}}+C_{Q_{1}}^{\ast }C_{9}^{eff})  \notag \\
&&+\frac{s}{2m_{\Lambda _{b}}}\bigg((f_{2}-g_{2}m_{\Lambda
})^{2}-g_{2}^{2}m_{\Lambda _{b}}^{2})(C_{9}^{\ast
eff}C_{10}+C_{10}^{\ast
}C_{9}^{eff})  \notag \\
&&+m_{b}(f_{2}^{2}-g_{2}^{2}s)(C_{10}^{\ast }C_{7}^{eff}+C_{7}^{\ast
eff}C_{10})\bigg)(2m_{l}^{2}+m_{\Lambda }^{2}+m_{\Lambda
_{b}}^{2}-s-u_{\max }-u_{\min })  \notag \\
&&+\frac{m_{l}}{2m_{b}}f_{2}^{2}(m_{\Lambda }-m_{\Lambda _{b}})((m_{\Lambda
}+m_{\Lambda _{b}})^{2}-s)(C_{10}^{\ast }C_{Q_{2}}+C_{Q_{2}}^{\ast }C_{10})
\notag \\
&&-\frac{m_{b}}{s}(2m_{l}^{2}+s)((m_{\Lambda }^{2}-m_{\Lambda
_{b}}^{2}+s)f_{2}^{2}-4sm_{\Lambda }f_{2}g_{2}+g_{2}^{2}s(m_{\Lambda
}^{2}-m_{\Lambda _{b}}^{2}+s))(C_{9}^{\ast eff}C_{7}^{eff}+C_{7}^{\ast
eff}C_{9}^{eff})  \notag \\
&&+\frac{sm_{\Lambda _{b}}}{2m_{b}^{2}}f_{2}^{2}((m_{\Lambda }+m_{\Lambda
_{b}})^{2}-s)(C_{Q_{2}}C_{Q_{2}}^{\ast })+\frac{2m_{b}^{2}}{3m_{\Lambda
_{b}}s}\bigg((6m_{l}^{2}(sm_{l}^{2}+(m_{\Lambda }^{2}-m_{\Lambda
_{b}}^{2})^{2}-s(s+u_{\max }+u_{\min }))  \notag \\
&&+s(3m_{\Lambda }^{4}+3m_{\Lambda _{b}}^{4}-3(m_{\Lambda }^{2}+m_{\Lambda
_{b}}^{2})(s+u_{\max }+u_{\min })+2u_{\max }^{2}  \notag \\
&&+3s(u_{\max }+u_{\min })+2u_{\max }u_{\min
}))f_{2}^{2}+6f_{2}g_{2}m_{\Lambda }(m_{\Lambda }^{2}-m_{\Lambda
_{b}}^{2}-s)s(2m_{l}^{2}+s)  \notag \\
&&+g_{2}^{2}s^{2}(-6m_{l}^{4}+6(u_{\max }+u_{\min
})m_{l}^{2}-3s^{2}-2(u_{\max }^{2}-u_{\min }^{2})  \notag \\
&&-3s(u_{\max }+u_{\min })+3m_{\Lambda _{b}}^{2}(s+u_{\max }+u_{\min
})-2u_{\max }u_{\min }  \notag \\
&&+3m_{\Lambda }^{2}(-2m_{\Lambda _{b}}^{2}+s+u_{\max }+u_{\min }))\bigg)%
\left| C_{7}^{eff}\right| ^{2}  \notag \\
&&+\frac{1}{4m_{\Lambda _{b}}}\bigg((-6m_{l}^{4}+(6m_{l}^{2}-3s)(u_{\max
}+u_{\min })-3s^{2}-2(u_{\max }^{2}+u_{\min }^{2}+u_{\max }u_{\min
})+3m_{\Lambda _{b}}^{2}(s+u_{\max }+u_{\min })  \notag \\
&&+3m_{\Lambda }^{2}(-2m_{\Lambda _{b}}^{2}+s+u_{\max }+u_{\min
}))f_{2}^{2}-6f_{2}g_{2}m_{\Lambda }(m_{\Lambda }^{2}-m_{\Lambda
_{b}}^{2}-s)s(2m_{l}^{2}+s)  \notag \\
&&+g_{2}^{2}(6sm_{l}^{4}+6m_{l}^{2}((m_{\Lambda }^{2}-m_{\Lambda
_{b}}^{2})^{2}-s(s+u_{\max }+u_{\min }))+s(3m_{\Lambda }^{4}+3m_{\Lambda
_{b}}^{4}  \notag \\
&&-3(m_{\Lambda }^{2}+m_{\Lambda _{b}}^{2})(s+u_{\max }+u_{\min })+2(u_{\max
}^{2}+u_{\min }^{2}+u_{\max }u_{\min })+3t(u_{\max }+u_{\min }))))\bigg)%
\left| C_{9}^{eff}\right| ^{2}  \notag \\
&&\frac{1}{4m_{\Lambda _{b}}}\bigg((6m_{l}^{4}+6(2m_{\Lambda
}^{2}+2m_{\Lambda _{b}}^{2}-2s-u_{\max }-u_{\min })m_{l}^{2}+2(u_{\max
}^{2}+u_{\min }^{2}+u_{\max }u_{\min })  \notag \\
&&+3(s-m_{\Lambda _{b}}^{2}-m_{\Lambda }^{2})(s+u_{\max }+u_{\min
}))f_{2}^{2}+6g_{2}f_{2}(4m_{l}^{2}-s)(m_{\Lambda _{b}}^{2}-m_{\Lambda
}^{2}+s)  \notag \\
&&+g_{2}^{2}(-6sm_{l}^{4}+6(m_{\Lambda }^{4}-(2m_{\Lambda
_{b}}^{2}+s)m_{\Lambda }^{2}+m_{\Lambda _{b}}^{4}-m_{\Lambda
_{b}}^{2}s+s(u_{\max }+u_{\min })m_{l}^{2}  \notag \\
&&+s(-3m_{\Lambda }^{4}+3(m_{\Lambda }^{2}+m_{\Lambda _{b}}^{2})(s+u_{\max
}+u_{\min })-3m_{\Lambda _{b}}^{4}  \notag \\
&&-2(u_{\max }^{2}+u_{\min }^{2}+u_{\max }u_{\min })-3t(u_{\max }+u_{\min
})))\bigg)\left| C_{10}\right| ^{2}\bigg\},  \notag \\
&&  \label{drate}
\end{eqnarray}
where $u_{max}$ and $u_{min}$ are defined in Eq.
(\ref{defination-u}). In Eq. (\ref{drate}) we have given the result
in MSSM with NHBs and ignored the contribution from the primed
operators which appear SUSY SO(10) GUT model as the results are very
tiny.

\subsection{FBAs of $\Lambda _{b}\rightarrow \Lambda
l^{+}l^{-}$}

Now we are in a position to explore the FBAs of $\Lambda
_{b}\rightarrow \Lambda l^{+}l^{-}$, which is an essential
observable sensitive to the new physics effects. To calculate the
forward-backward asymmetry, we consider the following double
differential decay rate formula for the process $\Lambda
_{b}\rightarrow \Lambda l^{+}l^{-}$
\begin{equation}
{\frac{d^{2}\Gamma (s,\cos \theta )}{dsd\cos \theta }}={\frac{1}{(2\pi )^{3}}%
}{\frac{1}{64m_{\Lambda _{b}}^{3}}}\lambda ^{1/2}(m_{\Lambda
_{b}}^{2},m_{\Lambda }^{2},s)\sqrt{1-{\frac{4m_{l}^{2}}{s}}}|{\widetilde{M}}%
_{\Lambda _{b}\rightarrow \Lambda l^{+}l^{-}}|^{2}\,\,,
\end{equation}%
where $\theta $ is the angle between the momentum of $\Lambda _{b}$
baryon and $l^{-}$ in the dilepton rest frame; $\lambda
(a,b,c)=a^{2}+b^{2}+c^{2}-2ab-2ac-2bc$. Following Refs. \cite{c.q.
geng 4, b to s in theory 9}, the differential and normalized FBAs
for the semi-leptonic decay $\Lambda _{b}\rightarrow \Lambda
l^{+}l^{-}$ are defined as
\begin{equation}
{\frac{dA_{FB}(q^{2})}{ds}}=\int_{0}^{1}d\cos \theta {\frac{d^{2}\Gamma
(s,\cos \theta )}{dsd\cos \theta }}-\int_{-1}^{0}d\cos \theta {\frac{%
d^{2}\Gamma (s,\cos \theta )}{dsd\cos \theta }}
\end{equation}%
and
\begin{equation}
A_{FB}(q^{2})={\frac{\int_{0}^{1}d\cos \theta {\frac{d^{2}\Gamma (s,\cos
\theta )}{dsd\cos \theta }}-\int_{-1}^{0}d\cos \theta {\frac{d^{2}\Gamma
(s,\cos \theta )}{dsd\cos \theta }}}{\int_{0}^{1}d\cos \theta {\frac{%
d^{2}\Gamma (s,\cos \theta )}{dsd\cos \theta }}+\int_{-1}^{0}d\cos \theta {%
\frac{d^{2}\Gamma (s,\cos \theta )}{dsd\cos \theta }}}}.
\end{equation}
Following the same procedure as we did for the differential decay rate, one
can easily get the expression for the forward-backward asymmetry.

\subsection{Lepton Polarization asymmetries of $\Lambda _{b}\rightarrow
\Lambda l^{+}l^{-}$}

In the rest frame of the lepton $l^{-}$, the unit vectors along
longitudinal, normal and transversal component of the $l^{-}$ can be
defined as \cite{Aliev UED}:%
\begin{eqnarray}
s_{L}^{-\mu } &=&(0,\vec{e}_{L})=\left( 0,\frac{\vec{p}_{-}}{\left| \vec{p}%
_{-}\right| }\right) ,  \notag \\
s_{N}^{-\mu } &=&(0,\vec{e}_{N})=\left( 0,\frac{\vec{p}_{\Lambda }\times
\vec{p}_{-}}{\left| \vec{p}_{\Lambda }\times \vec{p}_{-}\right| }\right) ,
\label{p-vectors} \\
s_{T}^{-\mu } &=&(0,\vec{e}_{T})=\left( 0,\vec{e}_{N}\times \vec{e}%
_{L}\right),  \notag
\end{eqnarray}%
where $\vec{p}_{-}$ and $\vec{p}_{\Lambda }$  are the three-momenta
of the lepton $l^{-}$ and $\Lambda $ baryon respectively in the center mass (CM) frame of $%
l^{+}l^{-}$ system. Lorentz transformation is used to boost the
longitudinal component of the lepton polarization to the CM frame of
the lepton pair as
\begin{equation}
\left( s_{L}^{-\mu }\right) _{CM}=\left( \frac{|\vec{p}_{-}|}{m_{l}},\frac{%
E_{l}\vec{p}_{-}}{m_{l}\left| \vec{p}_{-}\right| }\right)
\label{bossted component}
\end{equation}%
where $E_{l}$ and $m_{l}$ are the energy and mass of the lepton in the CM\
frame. The normal and transverse components remain unchanged under the
Lorentz boost.

The longitudinal ($P_{L}$), normal ($P_{N}$) and the transverse ($P_{T}$)
polarizations of lepton can be defined as:%
\begin{equation}
P_{i}^{(\mp )}(s)=\frac{\frac{d\Gamma }{ds}(\vec{\xi}^{\mp }=\vec{e}^{\mp })-%
\frac{d\Gamma }{ds}(\vec{\xi}^{\mp }=-\vec{e}^{\mp })}{\frac{d\Gamma }{ds}(%
\vec{\xi}^{\mp }=\vec{e}^{\mp })+\frac{d\Gamma }{ds}(\vec{\xi}^{\mp }=-\vec{e%
}^{\mp })}  \label{polarization-defination}
\end{equation}%
where $i=L,\;N,\;T$ and $\vec{\xi}^{\mp }$ is the spin direction along the
leptons $l^{\mp }$. The differential decay rate for polarized lepton $l^{\mp
}$ in $\Lambda _{b}\rightarrow \Lambda l^{+}l^{-}$ decay along any spin
direction $\vec{\xi}^{\mp }$ is related to the unpolarized decay rate (\ref%
{differential decay width}) with the following relation%
\begin{equation}
\frac{d\Gamma (\vec{\xi}^{\mp })}{ds}=\frac{1}{2}\left( \frac{d\Gamma }{ds}%
\right) [1+(P_{L}^{\mp }\vec{e}_{L}^{\mp }+P_{N}^{\mp }\vec{e}_{N}^{\mp
}+P_{T}^{\mp }\vec{e}_{T}^{\mp })\cdot \vec{\xi}^{\mp }].
\label{polarized-decay}
\end{equation}%
We can achieve the
expressions of longitudinal, normal and transverse polarizations for $%
\Lambda _{b}\rightarrow \Lambda l^{+}l^{-}$ decays as collected
below, where only the results in MSSM models with NHB are given for
the conciseness of paper. Thus the longitudinal lepton polarization
can be written as
\begin{eqnarray}
P_{L}(s) &=&(1/{\frac{d\Gamma }{ds}})\frac{\alpha ^{2}G_{F}^{2}\left|
V_{tb}V_{ts}^{\ast }\right| ^{2}u_{\max }}{768(m_{b}^{2}-m_{s}^{2})m_{%
\Lambda _{b}}^{3}\pi ^{5}}\sqrt{1-\frac{4m_{l}^{2}}{s}}\times   \notag \\
&&\bigg\{-6m_{l}(m_{b}-m_{s})m_{\Lambda _{b}}f_{2}^{2}(m_{\Lambda
}-m_{\Lambda _{b}})(s-(m_{\Lambda }+m_{\Lambda
_{b}})^{2})(C_{Q_{1}}C_{10}^{\ast }+C_{Q_{1}}^{\ast }C_{10})  \notag \\
&&-6m_{\Lambda _{b}}^{2}f_{2}s(s-(m_{\Lambda }+m_{\Lambda
_{b}})^{2})(C_{Q_{1}}^{\ast }C_{Q_{2}}+C_{Q_{1}}C_{Q_{2}}^{\ast })  \notag \\
&&+(m_{b}^{2}-m_{s}^{2})((3m_{\Lambda }^{4}+3m_{\Lambda
_{b}}^{4}-3s^{2})(f_{2}^{2}+g_{2}^{2}s)-6m_{\Lambda _{b}}^{2}m_{\Lambda
}^{2}(f_{2}^{2}+g_{2}^{2}s)  \notag \\
&&-\sqrt{\frac{\lambda (m_{\Lambda _{b}}^{2},m_{\Lambda }^{2},s)}{%
1-4m_{l}^{2}/s}}(f_{2}^{2}-g_{2}^{2}s))(C_{9}^{\ast
eff}C_{10}+C_{9}^{eff}C_{10}^{\ast })  \notag \\
&&-12(m_{b}^{2}-m_{s}^{2})(m_{b}m_{\Lambda _{b}}(m_{\Lambda }^{2}-m_{\Lambda
_{b}}^{2}+s)-m_{s}m_{\Lambda }(m_{\Lambda }^{2}-m_{\Lambda _{b}}^{2}-s)
\notag \\
&&(f_{2}^{2}+g_{2}^{2}s)+f_{2}g_{2}(m_{s}((m_{\Lambda }^{2}-m_{\Lambda
_{b}}^{2})^{2}-s)-4m_{b}m_{\Lambda _{b}}m_{\Lambda }s))(C_{10}^{\ast
}C_{7}^{eff}+C_{7}^{\ast eff}C_{10})\bigg\}.  \notag \\
&&  \label{expression-LP}
\end{eqnarray}

Similarly, the normal lepton polarization is
\begin{eqnarray}
P_{N}(s) &=&(1/{\frac{d\Gamma }{ds}})\frac{\alpha ^{2}G_{F}^{2}\left|
V_{tb}V_{ts}^{\ast }\right| ^{2}u_{\max }}{1024(m_{b}+m_{s})m_{\Lambda
_{b}}^{3}\pi ^{4}}\sqrt{\frac{\lambda (m_{\Lambda _{b}}^{2},m_{\Lambda
}^{2},s)}{s}}\times   \notag \\
&&\bigg\{(s-4m_{l}^{2})(g_{2}s-f_{2}(m_{\Lambda }+m_{\Lambda
_{b}}))m_{\Lambda _{b}}(C_{Q_{1}}C_{10}^{\ast }+C_{Q_{1}}^{\ast }C_{10})
\notag \\
&&-2sm_{\Lambda _{b}}\frac{(m_{b}+m_{s})^{2}}{m_{b}-m_{s}}(g_{2}(m_{\Lambda
}+m_{\Lambda _{b}})-f_{2})f_{2}(C_{Q_{2}}^{\ast }C_{7}^{eff}+C_{7}^{\ast
eff}C_{Q_{2}})  \notag \\
&&+s\frac{m_{b}+m_{s}}{m_{b}-m_{s}}m_{\Lambda _{b}}(f_{2}(m_{\Lambda
}+m_{\Lambda _{b}})-g_{2}s)f_{2}(C_{Q_{2}}^{\ast }C_{9}^{eff}+C_{9}^{\ast
eff}C_{Q_{2}})  \notag \\
&&-2f_{2}m_{l}(m_{b}+m_{s})(f_{2}(m_{\Lambda _{b}}^{2}-m_{\Lambda
}^{2})+g_{2}m_{\Lambda }s)(C_{9}^{eff}C_{10}^{\ast }+C_{9}^{\ast eff}C_{10})
\notag \\
&&+8m_{l}(m_{b}+m_{s})(m_{s}m_{\Lambda }(f_{2}^{2}+g_{2}^{2}s)  \notag \\
&&+m_{b}m_{\Lambda _{b}}(f_{2}^{2}-g_{2}^{2}s)-f_{2}g_{2}m_{s}(m_{\Lambda
}^{2}-m_{\Lambda _{b}}^{2}+s))(C_{9}^{eff}C_{7}^{\ast eff}+C_{9}^{\ast
eff}C_{7}^{eff})  \notag \\
&&+4m_{l}(m_{b}+m_{s})f_{2}(g_{2}m_{s}m_{\Lambda
_{b}}^{2}-f_{2}m_{b}m_{\Lambda _{b}}+m_{\Lambda }m_{s}(f_{2}-g_{2}m_{\Lambda
}))(C_{7}^{\ast eff}C_{10}+C_{7}^{eff}C_{10}^{\ast })  \notag \\
&&-\frac{16(m_{b}+m_{s})^{2}(m_{b}-m_{s})m_{l}}{s}(f_{2}(m_{\Lambda
_{b}}-m_{\Lambda })+g_{2}s)(g_{2}s-f_{2}(m_{\Lambda _{b}}-m_{\Lambda
}))\left| C_{7}^{eff}\right| ^{2}  \notag \\
&&+4m_{l}(m_{b}+m_{s})s((f_{2}-g_{2}m_{\Lambda
})^{2}-g_{2}^{2}m_{\Lambda _{b}}^{2})\left| C_{9}^{eff}\right|
^{2}\bigg\} , \label{expression-PN}
\end{eqnarray}%
and the transverse one is given by
\begin{eqnarray}
P_{T}(s) &=&(1/{\frac{d\Gamma }{ds}})\frac{i\alpha ^{2}G_{F}^{2}\left|
V_{tb}V_{ts}^{\ast }\right| ^{2}u_{\max }}{4096m_{\Lambda _{b}}^{3}\pi ^{4}%
\sqrt{s}}\sqrt{1-\frac{4m_{l}^{2}}{s}}\sqrt{\lambda (m_{\Lambda
_{b}}^{2},m_{\Lambda }^{2},s)}(m_{\Lambda _{b}}^{2}-m_{\Lambda
}^{2}+s)\times   \notag \\
&&\bigg\{2m_{\Lambda _{b}}f_{2}(g_{2}(m_{\Lambda }+m_{\Lambda
_{b}})-f_{2})(C_{Q_{1}}C_{7}^{\ast eff}-C_{Q_{1}}^{\ast }C_{7}^{eff})  \notag
\\
&&-\frac{m_{\Lambda _{b}}}{m_{b}+m_{s}}f_{2}(g_{2}s-f_{2}(m_{\Lambda
}+m_{\Lambda _{b}}))(C_{Q_{1}}C_{9}^{\ast eff}-C_{9}^{eff}C_{Q_{1}})  \notag
\\
&&-\frac{m_{\Lambda _{b}}}{m_{b}-m_{s}}f_{2}(g_{2}s-f_{2}(m_{\Lambda
}+m_{\Lambda _{b}}))(C_{Q_{2}}C_{10}^{\ast }-C_{10}C_{Q_{2}}^{\ast })  \notag
\\
&&-2m_{l}((f_{2}-g_{2}m_{\Lambda })^{2}-g_{2}^{2}m_{\Lambda
_{b}}^{2})(C_{9}^{\ast eff}C_{10}-C_{9}^{eff}C_{10}^{\ast })  \notag \\
&&+\frac{4m_{l}}{s}((m_{s}m_{\Lambda }+m_{b}m_{\Lambda
_{b}})f_{2}^{2}-f_{2}g_{2}m_{s}(m_{\Lambda }^{2}-m_{\Lambda _{b}}^{2}+s)
\notag \\
&&+g_{2}^{2}(m_{s}m_{\Lambda }-m_{b}m_{\Lambda _{b}})s)(C_{10}^{\ast
}C_{7}^{eff}-C_{7}^{\ast eff}C_{10})\bigg\}. \notag \\
&&  \label{expression-TP}
\end{eqnarray}%
The ${\frac{d\Gamma }{ds}}$ appearing in the above equation is the
one given in Eq. (\ref{drate}).

\subsection{$\Lambda $ polarization in $\Lambda _{b}\rightarrow \Lambda
l^{+}l^{-}$}

To study the $\Lambda $ spin polarization, one needs to express the
$\Lambda $ four spin vector in terms of a unit vector $\hat{\xi}$
along the $\Lambda $
spin in its rest frame as \cite{c.q. geng 2}%
\begin{equation}
s_{0}=\frac{\vec{p}_{\Lambda }\cdot \vec{\xi}}{m_{\Lambda }},\text{ }\vec{s}=%
\vec{\xi}+\frac{s_{0}}{E_{\Lambda }+m_{\Lambda }}\vec{p}_{\Lambda },
\label{polarization-lambda}
\end{equation}%
where the unit vectors along the longitudinal, normal and transverse
components of the $\Lambda $ polarization are chosen to be
\begin{eqnarray*}
\hat{e}_{L} &=&\frac{\vec{p}_{\Lambda }}{\left| \vec{p}_{\Lambda }\right| },
\\
\hat{e}_{N} &=&\frac{\vec{p}_{\Lambda }\times (\vec{p}_{-}\times \vec{p}%
_{\Lambda })}{\left| \vec{p}_{\Lambda }\times (\vec{p}_{-}\times \vec{p}%
_{\Lambda })\right| }, \\
\hat{e}_{T} &=&\frac{\vec{p}_{-}\times \vec{p}_{\Lambda }}{\left| \vec{p}%
_{-}\times \vec{p}_{\Lambda }\right| }.
\end{eqnarray*}%
Similar to the lepton polarization, the polarization asymmetries for $%
\Lambda $ baryon in $\Lambda _{b}\rightarrow \Lambda l^{+}l^{-}$ can be
defined as
\begin{equation}
P_{i}^{(\mp )}(s)=\frac{\frac{d\Gamma }{ds}(\vec{\xi}=\hat{e})-\frac{d\Gamma
}{ds}(\vec{\xi}=-\hat{e})}{\frac{d\Gamma }{ds}(\vec{\xi}=\hat{e})+\frac{%
d\Gamma }{ds}(\vec{\xi}=-\hat{e})}  \label{LNT for Lambda}
\end{equation}%
where $i=L,\;N,\;T$ and $\vec{\xi}$ is the spin direction along the $\Lambda
$ baryon. The differential decay rate for polarized $\Lambda $ baryon in $%
\Lambda _{b}\rightarrow \Lambda l^{+}l^{-}$ decay along any spin direction $%
\vec{\xi} $ is related to the unpolarized decay rate (\ref{differential
decay width}) through the following relation%
\begin{equation}
\frac{d\Gamma (\vec{\xi})}{ds}=\frac{1}{2}\left( \frac{d\Gamma }{ds}\right)
[1+(P_{L}\vec{e}_{L}+P_{N}\vec{e}_{N}+P_{T}\vec{e}_{T})\cdot \vec{\xi}].
\label{decay rate polarized Lambda}
\end{equation}%
Following the same procedure as we did for the lepton polarizations,
we can derive the formulae for the longitudinal, normal and
transverse polarizations of $\Lambda $ baryon in the MSSM as
\begin{eqnarray}
P_{L}(s) &=&(1/{\frac{d\Gamma }{ds}})\frac{\alpha ^{2}G_{F}^{2}\left|
V_{tb}V_{ts}^{\ast }\right| ^{2}u_{\max }}{64m_{\Lambda }m_{\Lambda
_{b}}^{3}\pi ^{5}s^{3/2}}\times  \notag \\
&&\bigg\{\frac{m_{\Lambda }m_{\Lambda _{b}}m_{l}}{2m_{b}}(m_{\Lambda
}+m_{\Lambda _{b}})s^{3/2}\sqrt{\lambda _{p}}f_{2}^{2}(C_{Q_{2}}C_{10}^{\ast
}+C_{10}^{\ast }C_{Q_{2}})  \notag \\
&&+m_{\Lambda }m_{\Lambda _{b}}m_{l}(2m_{l}^{2}+s)(g_{2}^{2}s-f_{2}^{2})%
\sqrt{\lambda _{p}}\sqrt{s}(C_{7}^{\ast
eff}C_{9}^{eff}+C_{7}^{eff}C_{9}^{\ast eff})  \notag \\
&&+\frac{m_{b}^{2}}{3}\bigg[\frac{m_{\Lambda }}{\sqrt{s}}(12m_{l}^{2}(m_{%
\Lambda }^{2}-m_{\Lambda _{b}}^{2})\sqrt{\lambda _{p}}+s\sqrt{1-\frac{%
4m_{l}^{2}}{s}}u_{\max }(m_{\Lambda }^{2}-m_{\Lambda _{b}}^{2}+s)-3\sqrt{%
\lambda _{p}}s(s-m_{\Lambda }^{2}+m_{\Lambda _{b}}^{2}))f_{2}^{2}  \notag \\
&&-g_{2}f_{2}(24m_{l}^{2}\sqrt{s}\sqrt{\lambda _{p}}m_{\Lambda }^{2}-\sqrt{s}%
u_{\max }^{2}\sqrt{\lambda _{p}}+\sqrt{s}u_{\max }\sqrt{1-\frac{4m_{l}^{2}}{s%
}}\lambda _{p})  \notag \\
&&+g_{2}^{2}m_{\Lambda }s^{3/2}(12\sqrt{\lambda _{p}}m_{l}^{2}+u_{\max }%
\sqrt{1-\frac{4m_{l}^{2}}{s}}(m_{\Lambda }^{2}-m_{\Lambda _{b}}^{2}+s)+3%
\sqrt{\lambda _{p}}(s-m_{\Lambda }^{2}+m_{\Lambda _{b}}^{2}))\bigg]\left|
C_{7}^{eff}\right| ^{2}  \notag \\
&&+\frac{s}{12}\bigg[-m_{\Lambda }(-12m_{l}^{2}\sqrt{s}\sqrt{\lambda _{p}}+%
\sqrt{s}\sqrt{1-\frac{4m_{l}^{2}}{s}}u_{\max }(m_{\Lambda }^{2}-m_{\Lambda
_{b}}^{2}+s)+3u_{\max }\sqrt{s}(s-m_{\Lambda }^{2}+m_{\Lambda
_{b}}^{2}))f_{2}^{2}  \notag \\
&&+g_{2}f_{2}u_{\max }\sqrt{s}(\sqrt{1-\frac{4m_{l}^{2}}{s}}\lambda
_{p}-u_{\max }\sqrt{\lambda _{p}})-g_{2}^{2}m_{\Lambda }\sqrt{s}(12m_{l}^{2}%
\sqrt{\lambda _{p}}(s-m_{\Lambda }^{2}+m_{\Lambda _{b}}^{2})  \notag \\
&&+u_{\max }s\sqrt{1-\frac{4m_{l}^{2}}{s}}(m_{\Lambda }^{2}-m_{\Lambda
_{b}}^{2}+s)-3s\sqrt{\lambda _{p}}(s-m_{\Lambda }^{2}+m_{\Lambda _{b}}^{2}))%
\bigg]\left| C_{10}\right| ^{2}  \notag \\
&&+\frac{s}{12}\bigg[-m_{\Lambda }(12m_{l}^{2}\sqrt{s}\sqrt{\lambda _{p}}+%
\sqrt{s}\sqrt{1-\frac{4m_{l}^{2}}{s}}u_{\max }(m_{\Lambda }^{2}-m_{\Lambda
_{b}}^{2}+s)-3u_{\max }\sqrt{s}(m_{\Lambda _{b}}^{2}-s+m_{\Lambda
}^{2}))f_{2}^{2}  \notag \\
&&+g_{2}f_{2}u_{\max }\sqrt{s}(\sqrt{1-\frac{4m_{l}^{2}}{s}}\lambda
_{p}-u_{\max }\sqrt{\lambda _{p}})-g_{2}^{2}m_{\Lambda }\sqrt{s}(-12m_{l}^{2}%
\sqrt{\lambda _{p}}(m_{\Lambda }^{2}-m_{\Lambda _{b}}^{2})  \notag \\
&&+u_{\max }s\sqrt{1-\frac{4m_{l}^{2}}{s}}(m_{\Lambda _{b}}^{2}-m_{\Lambda
}^{2}-s)-3s\sqrt{\lambda _{p}}(m_{\Lambda }^{2}-s-m_{\Lambda _{b}}^{2}))%
\bigg]\left| C_{9}^{eff}\right| ^{2}\bigg\} ,  \notag
\label{polarizationlambda-LP} \\
P_{N}(s) &=&(1/{\frac{d\Gamma }{ds}})\frac{\alpha ^{2}G_{F}^{2}\left|
V_{tb}V_{ts}^{\ast }\right| ^{2}u_{\max }}{512m_{b}m_{\Lambda _{b}}^{3}\pi
^{4}\sqrt{s}}\sqrt{1-\frac{4m_{l}^{2}}{s}}\times  \notag \\
&&\bigg\{2m_{b}m_{l}m_{\Lambda _{b}}(s-(m_{\Lambda }+m_{\Lambda
_{b}})^{2})(f_{2}(m_{\Lambda _{b}}-m_{\Lambda
})+g_{2}s)(C_{Q_{1}}C_{7}^{\ast eff}+C_{Q_{1}}^{\ast }C_{7}^{eff})  \notag \\
&&+m_{l}m_{\Lambda _{b}}s(s-(m_{\Lambda }+m_{\Lambda
_{b}})^{2})(f_{2}+g_{2}(m_{\Lambda _{b}}-m_{\Lambda }))(C_{Q_{1}}C_{9}^{\ast
eff}+C_{Q_{1}}^{\ast }C_{9}^{eff})  \notag \\
&&-m_{b}s(m_{\Lambda _{b}}^{2}-m_{\Lambda }^{2}+s)(m_{\Lambda
}f_{2}^{2}-g_{2}(m_{\Lambda }^{2}-m_{\Lambda
_{b}}^{2}+s)f_{2}+g_{2}m_{\Lambda }t)(C_{10}C_{9}^{\ast eff}+C_{10}^{\ast
}C_{9}^{eff})  \notag \\
&&+4m_{b}m_{\Lambda _{b}}s(-m_{\Lambda }f_{2}^{2}+g_{2}(m_{\Lambda
}^{2}-m_{\Lambda _{b}}^{2}+s)f_{2}-g_{2}^{2}m_{\Lambda }t)(C_{10}C_{7}^{\ast
eff}+C_{10}^{\ast }C_{7}^{eff})\bigg\} ,  \notag
\label{lambdapolarization-PN} \\
P_{T}(s) &=&(1/{\frac{d\Gamma }{ds}})\frac{i\alpha ^{2}G_{F}^{2}\left|
V_{tb}V_{ts}^{\ast }\right| ^{2}u_{\max }\lambda _{p}^{1/2}}{%
512m_{b}m_{\Lambda _{b}}^{2}\pi ^{4}\sqrt{s}}m_{l.}\sqrt{1-\frac{4m_{l}^{2}}{%
s}}\times \bigg\{2m_{b}(f_{2}(m_{\Lambda }+m_{\Lambda
_{b}})-g_{2}s)(C_{Q_{1}}C_{7}^{\ast eff}-C_{Q_{1}}^{\ast }C_{7}^{eff})
\notag \\
&&  \notag \\
&&+s(g_{2}(m_{\Lambda }+m_{\Lambda _{b}})-f_{2})(C_{Q_{1}}C_{9}^{\ast
eff}-C_{Q_{1}}^{\ast }C_{9}^{eff})\bigg\},  \label{Lambdapolarization-PT}
\end{eqnarray}
where $\lambda_p = \lambda(m_{\Lambda_b}^2, m_{\Lambda}^2,s)$ and
the mass of strange quark is neglected to make the expressions more
compact.

\section{Numerical Analysis}

In this section, we would like to present the numerical analysis of
decay rates, FBAs and polarization asymmetries of the lepton and
$\Lambda$ baryon. The numerical values of Wilson coefficients and
other input parameters used in our analysis are borrowed from Ref. \cite%
{Yan, Li, YuMing} and collected in Tables II, III and IV.
\begin{table}[tbh]
\caption{{}Values of input parameters used in our numerical analysis}
\label{Input parameters}%
\begin{tabular}{cc}
\hline\hline
$G_{F}=1.166\times 10^{-2}$ GeV$^{-2}$ & $\left| V_{ts}\right|
=41.61_{-0.80}^{+0.10}\times 10^{-3}$ \\
$\left| {V_{tb}}\right| =0.9991$ & $m_{b}=\left( 4.68\pm 0.03\right) $ GeV
\\
{$m_{c}\left( m_{c}\right) =1.275_{-0.015}^{+0.015}$ GeV} & $m_{s}\left( 1%
\text{ GeV}\right) =\left( 142\pm 28\right) $ MeV \\ \hline
$m_{\Lambda _{b}}=5.62$ GeV & $m_{\Lambda }=1.12$ GeV \\ \hline
$f_{\Lambda _{b}}=3.9_{-0.2}^{+0.4}\times 10^{-3}$ GeV$^{2}$ & $f_{\Lambda
}=6.0_{-0.4}^{+0.4}\times 10^{-3}$ GeV$^{2}$ \\ \hline\hline
\end{tabular}%
\end{table}
\begin{table}[tbh]
\caption{{}Wilson Coefficients in SM\ and different SUSY models but
without Neutral Higgs boson contributions. The primed Wilson
coefficients corresponds to the operators which are opposite in
helicities from those of
the SM operators and these comes only in SUSY SO(10) GUT model.}%
\begin{tabular}{ccccccc}
\hline
Wilson Coefficients & $C_{7}^{eff}$ & $C_{7}^{\prime eff}$ & $C_{9}$ & $%
C_{9}^{\prime }$ & $C_{10}$ & $C_{10}^{\prime }$ \\ \hline
SM & $-0.313$ & $0$ & $4.334$ & $0$ & $-4.669$ & $0$ \\ \hline
SUSYI & $+0.3756$ & $0$ & $4.7674$ & $0$ & $-3.7354$ & $0$ \\ \hline
SUSYII & $+0.3756$ & $0$ & $4.7674$ & $0$ & $-3.7354$ & $0$ \\ \hline\hline
SUSYIII & $-0.3756$ & $0$ & $4.7674$ & $0$ & $-3.7354$ & $0$ \\ \hline\hline
SUSY SO(10) $\left( A_{0}=-1000\right) $ & $-0.219+0i$ & $0.039-0.038i$ & $%
4.275+0i$ & $0.011+0.0721i$ & $-4.732-0i$ & $-0.075-0.670i$ \\ \hline\hline
\end{tabular}%
\end{table}
\begin{table}[tbh]
\caption{{}Wilson coefficient corresponding to NHBs contributions.
SUSYI corresponds to the regions where SUSY can destructively
contribute and can change the sign of $C_{7}$, but contribution of
NHBs are neglected, SUSYII refers to the region where $\tan
\protect\beta $ is large and the masses of the superpartners are
relatively small. SUSY III corresponds to the regions where $\tan
\protect\beta $ is large and the masses of superpartners are
relatively large. The primed Wilson coefficients are the
contribution of NHBs in SUSY SO(10) GUT model. As the neutral Higgs
bosons are proportional to the lepton mass, and the values shown in
the table are for $\protect\mu $
and $\protect\tau $ case. The values in the bracket are for the $\protect%
\tau $.}
\label{NHB}%
\begin{tabular}{ccccc}
\hline
Wilson Coefficients & $C_{Q_{1}}$ & $C_{Q_{1}}^{\prime }$ & $C_{Q_{2}}$ & $%
C_{Q_{2}}^{\prime }$ \\ \hline
SM & $0$ & $0$ & $0$ & $0$ \\ \hline
SUSYI & $0$ & $0$ & 0 & $0$ \\ \hline
SUSYII & $6.5\left( 16.5\right) $ & $0$ & $-6.5\left( -16.5\right) $ & $0$
\\ \hline\hline
SUSYIII & $1.2\left( 4.5\right) $ & $0$ & $-1.2\left( -4.5\right) $ & $0$ \\
\hline\hline
SUSY SO(10) $\left( A_{0}=-1000\right) $ & $%
\begin{array}{c}
0.106+0i \\
\left( 1.775+0.002i\right)%
\end{array}%
$ & $%
\begin{array}{c}
-0.247+0.242i \\
\left( -4.148+4.074i\right)%
\end{array}%
$ & $%
\begin{array}{c}
-0.107+0i \\
\left( -1.797-0.002i\right)%
\end{array}%
$ & $%
\begin{array}{c}
-0.250+0.246i \\
\left( -4.202+4.128i\right)%
\end{array}%
$ \\ \hline\hline
\end{tabular}%
\end{table}
In the subsequent analysis, we will focus on the parameter space of large $%
\tan \beta $, where the NHBs effects are significant owing to the fact that
the Wilson coefficients corresponding to NHBs are proportional to $%
(m_{b}m_{l}/m_{h}) \tan ^{3} \beta $ $(h=h^{0}$, $A^{0})$. Here, one
$\tan \beta $ comes from the chargino-up-type squark loop and $\tan
^{2}\beta $
comes from the exchange of the NHBs. At large value of $\tan \beta $ the $%
C_{Q_{i}}^{(^{\prime })}$ compete with $C_{i}^{(^{\prime })}$ and
can overwhelm $C_{i}^{(^{\prime })}$ in some region as can be seen
from the Tables III and IV \cite{Huang}. Apart from the large $\tan
\beta $ limit, the other two conditions responsible for the large
contributions from NHBs are: (i) the mass values of the lighter
chargino and lighter stop should not be too large; (ii) the mass
splitting of charginos and stops should be large,
which also indicate large mixing between stop sector and chargino sector %
\cite{Yan}. Once these conditions are satisfied, the process
$B\rightarrow X_{s}\gamma $ will not only impose constraints on
$C_{7}$ but it also puts very stringent constraint on the possible
new physics. It is well known that the SUSY contribution is
sensitive to the sign of the Higgs mass term $\mu $ and SUSY
contributes destructively when the sign of this term becomes minus.
It is pointed out in literature \cite{Yan} that there exist
considerable regions of SUSY parameter space in which NHBs can
largely contribute to the process $b\rightarrow sl^{+}l^{-}$ due to
change of the sign of $C_{7}$ from positive to negative, while the
constraint on $b\rightarrow s\gamma $ is respected. Also, when the
masses of SUSY particles are relatively large, say about $450$ GeV,
there exist significant regions in the parameter space of SUSY
models in which NHBs could contribute largely. However, in these
cases $C_{7}$ does not change its sign, because contributions of
charged Higgs and charginos cancel each other. Hopefully, we can
distinguish between these two regions of SUSY by observing $\Lambda
_{b}\rightarrow \Lambda l^{+}l^{-}$ with ($l=\mu, \tau$).

\begin{figure}[tbp]
\begin{center}
\begin{tabular}{ccc}
\vspace{-2cm} \includegraphics[scale=0.6]{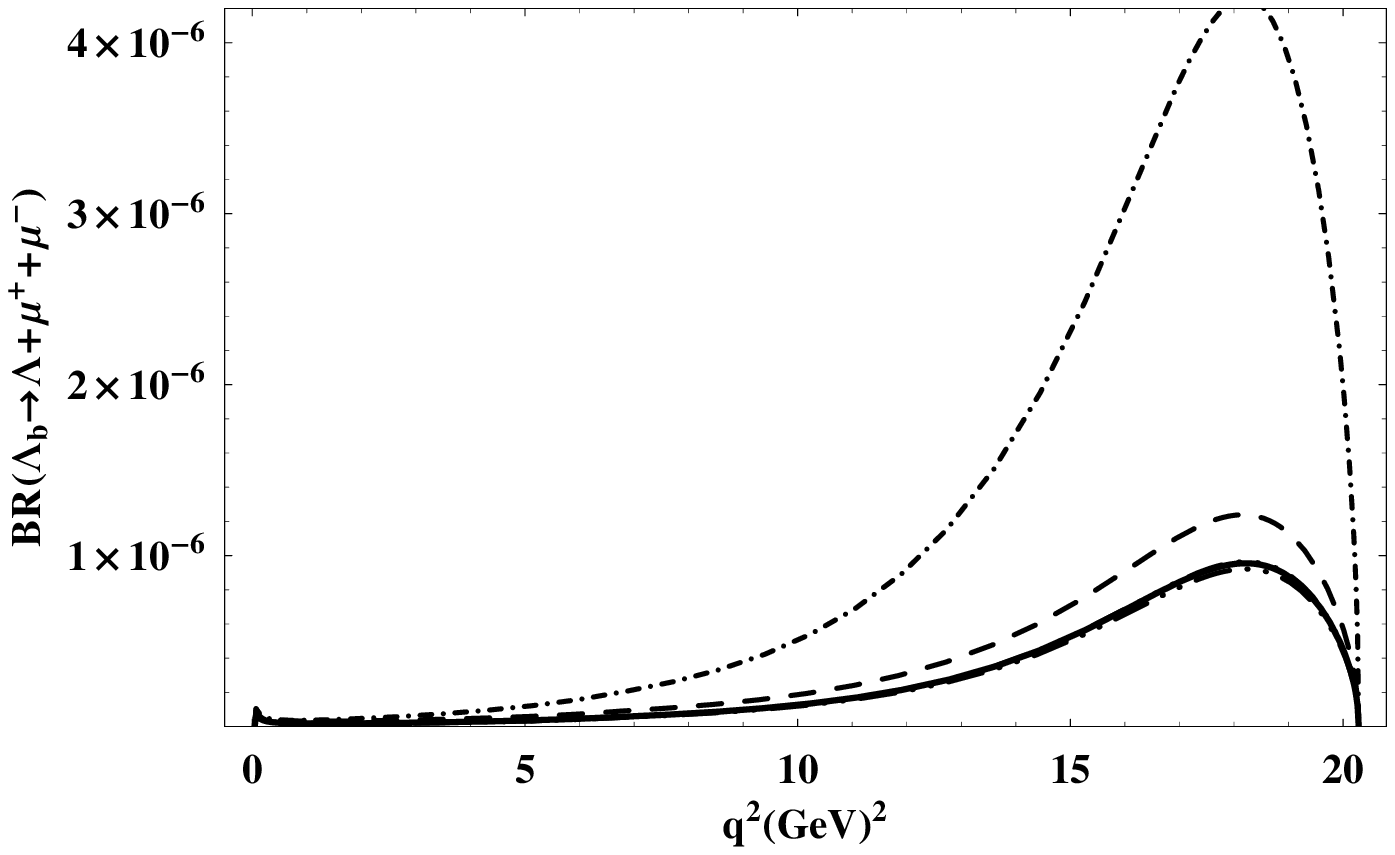} %
\includegraphics[scale=0.6]{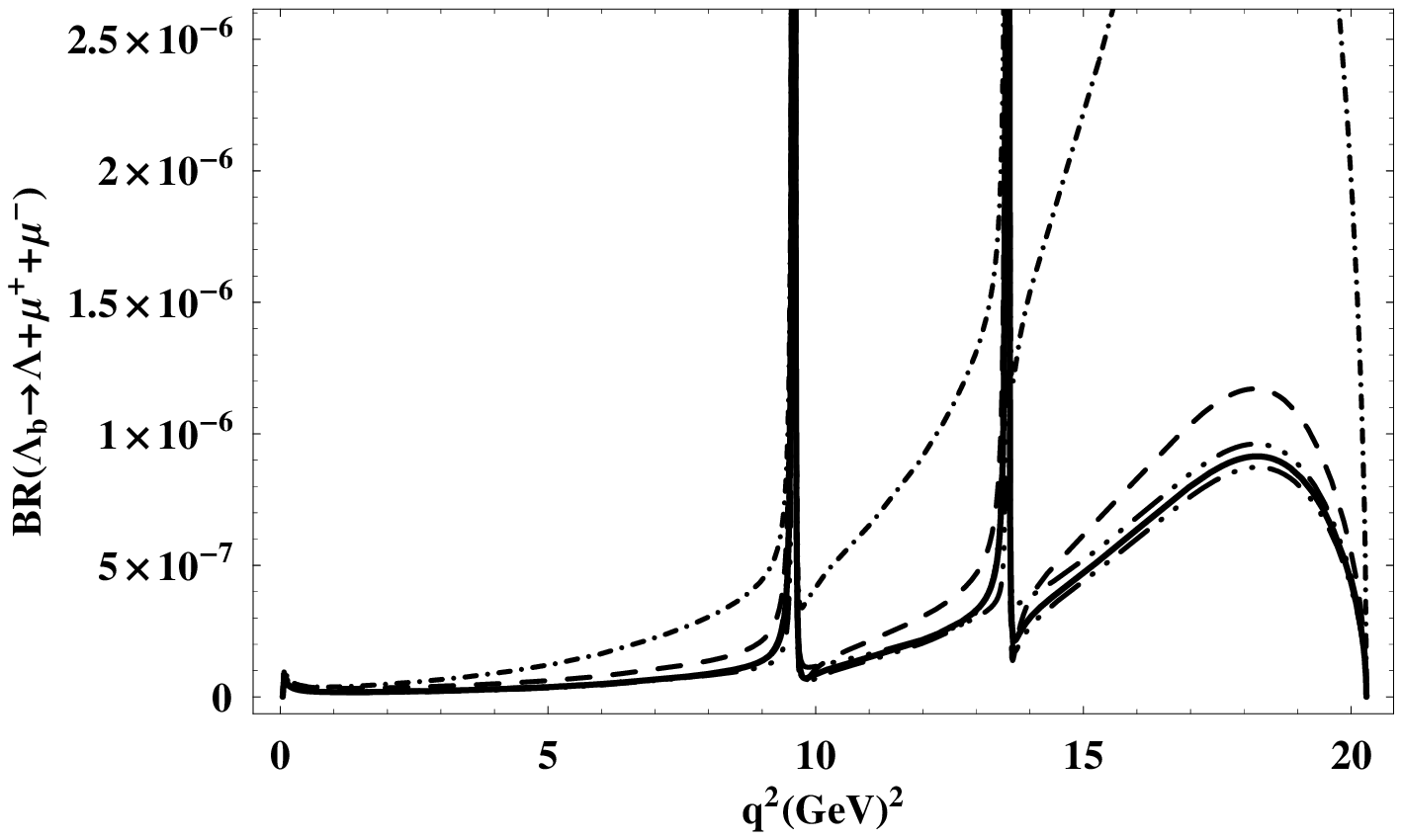} &  &  \\
\includegraphics[scale=0.6]{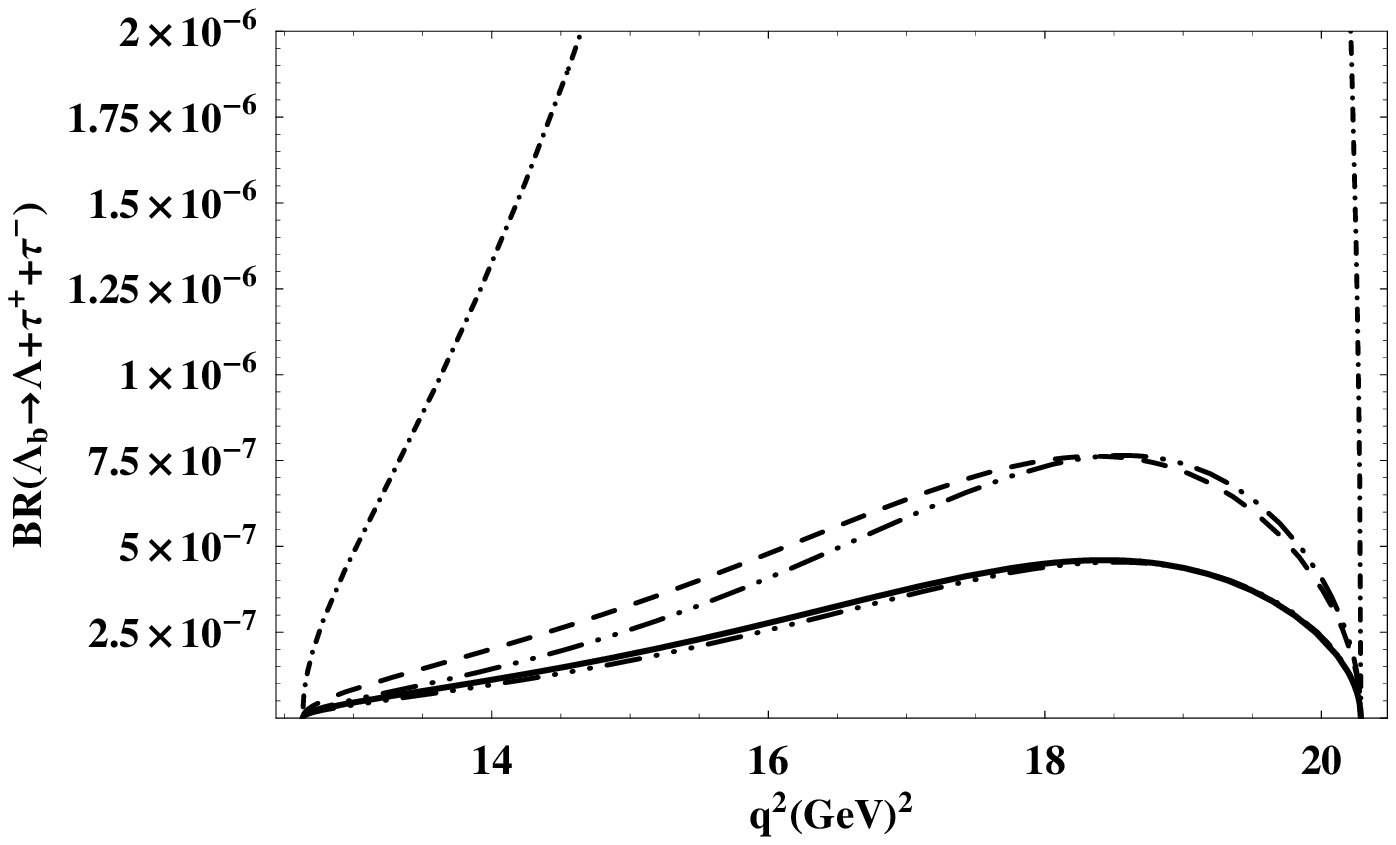} \includegraphics[scale=0.6]{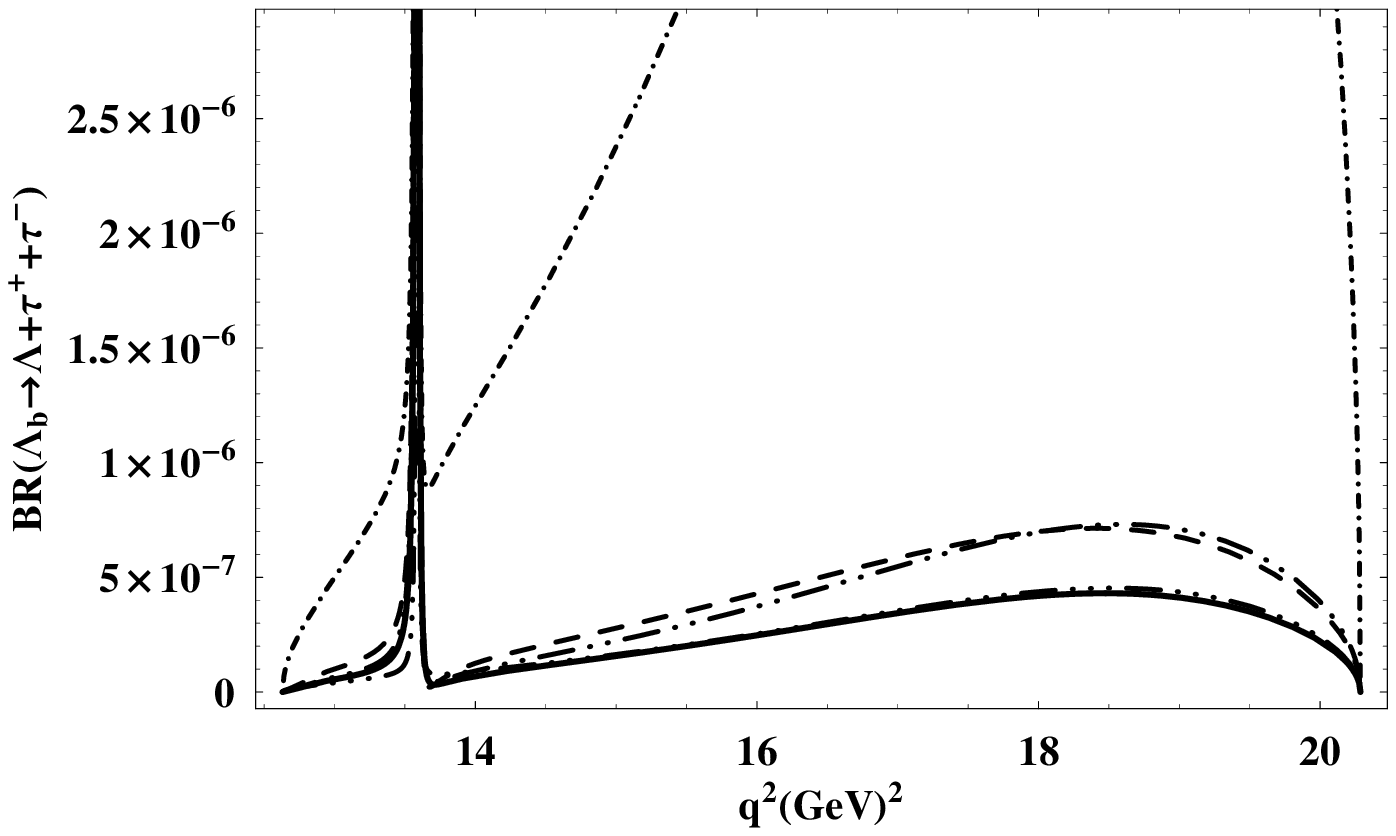}
\put (-350,220){(a)} \put (-100,220){(b)} \put (-350,30){(c)}
\put(-100,30){(d)} \vspace{-1cm} &  &
\end{tabular}%
\end{center}
\caption{The differential width for the $\Lambda_b \to \Lambda l^+l^-$ ($l=%
\protect\mu, \protect\tau$) decays as functions of $q^2$ without
long-distance contributions (a, c) and with long-distance contributions (b,
d). The solid, dashed, dashed-dot, dashed-double dot and dashed-triple dot
line represents, SM, SUSY I, SUSY II, SUSY III and SUSY SO(10) GUT model. }
\label{decay rate}
\end{figure}

The numerical results for the decay rates, FBAs and polarization
asymmetries of the lepton and $\Lambda $ baryon are presented in
Figs. 1-8. Fig. 1 describes the differential decay rate of $\Lambda
_{b}\rightarrow \Lambda l^{+}l^{-}$, from which one can see that the
supersymmetric effects are quite significant for SUSY I and SUSY II
model in the high momentum transfer regions for the final states
being muon pair, whereas these effects are extremely small for SUSY
III and SUSY SO(10) GUT models in this case. The reason for the
increase of differential decay width in SUSY I model is the relative
change in the sign of $C_{7}^{eff}$; while the large change in SUSY
II model is due to the contribution of the NHBs. As for the SUSY III
and SUSY SO (10) models, the value of the Wilson coefficients
corresponding to NHBs is small and hence one expects small
deviations from SM. For the tauon case, the values of Wilson
coefficients corresponding to NHBs in SUSY III are larger than that
for the muon case and therefore their effects are quite significant
as shown in Fig. 1(c, d). The numerical values of the branching
fractions for $\Lambda _{b}\rightarrow \Lambda l^{+}l^{-}$ ($l=\mu
,\tau $) with and without long-distance contribution in SM and
different SUSY models are given in Table V.
\begin{table}[tbh]
\caption{{}Branching ratio for $\Lambda _{b}\rightarrow \Lambda
l^{+}l^{-}\left( l=\protect\mu ,\protect\tau \right) $ in units of $10^{-6}$
in SM and different SUSY models.}%
\begin{tabular}{ccccc}
\hline
Branching Ratio & $%
\begin{array}{c}
\Lambda _{b}\rightarrow \Lambda \mu ^{+}\mu ^{-} \\
\text{without LD}%
\end{array}%
$ & $%
\begin{array}{c}
\Lambda _{b}\rightarrow \Lambda \mu ^{+}\mu ^{-} \\
\text{with LD}%
\end{array}%
$ & $%
\begin{array}{c}
\Lambda _{b}\rightarrow \Lambda \tau ^{+}\tau ^{-} \\
\text{without LD}%
\end{array}%
$ & $%
\begin{array}{c}
\Lambda _{b}\rightarrow \Lambda \tau ^{+}\tau ^{-} \\
\text{with LD}%
\end{array}%
$ \\ \hline
SM & $5.9$ & $39$ & $2.1$ & $4$ \\ \hline\hline
SUSYI & $7.9$ & $47$ & $3.5$ & $5.7$ \\ \hline\hline
SUSYII & $25$ & $65$ & $31$ & $33$ \\ \hline\hline
SUSYIII & $5.6$ & $45$ & $3.2$ & $5.6$ \\ \hline\hline
SUSY SO(10) $\left( A_{0}=-1000\right) $ & $5.92$ & $23$ & $2$ & $2.6$ \\
\hline\hline
\end{tabular}%
\end{table}

\begin{figure}[tbp]
\begin{center}
\begin{tabular}{ccc}
\vspace{-2cm} \includegraphics[scale=0.6]{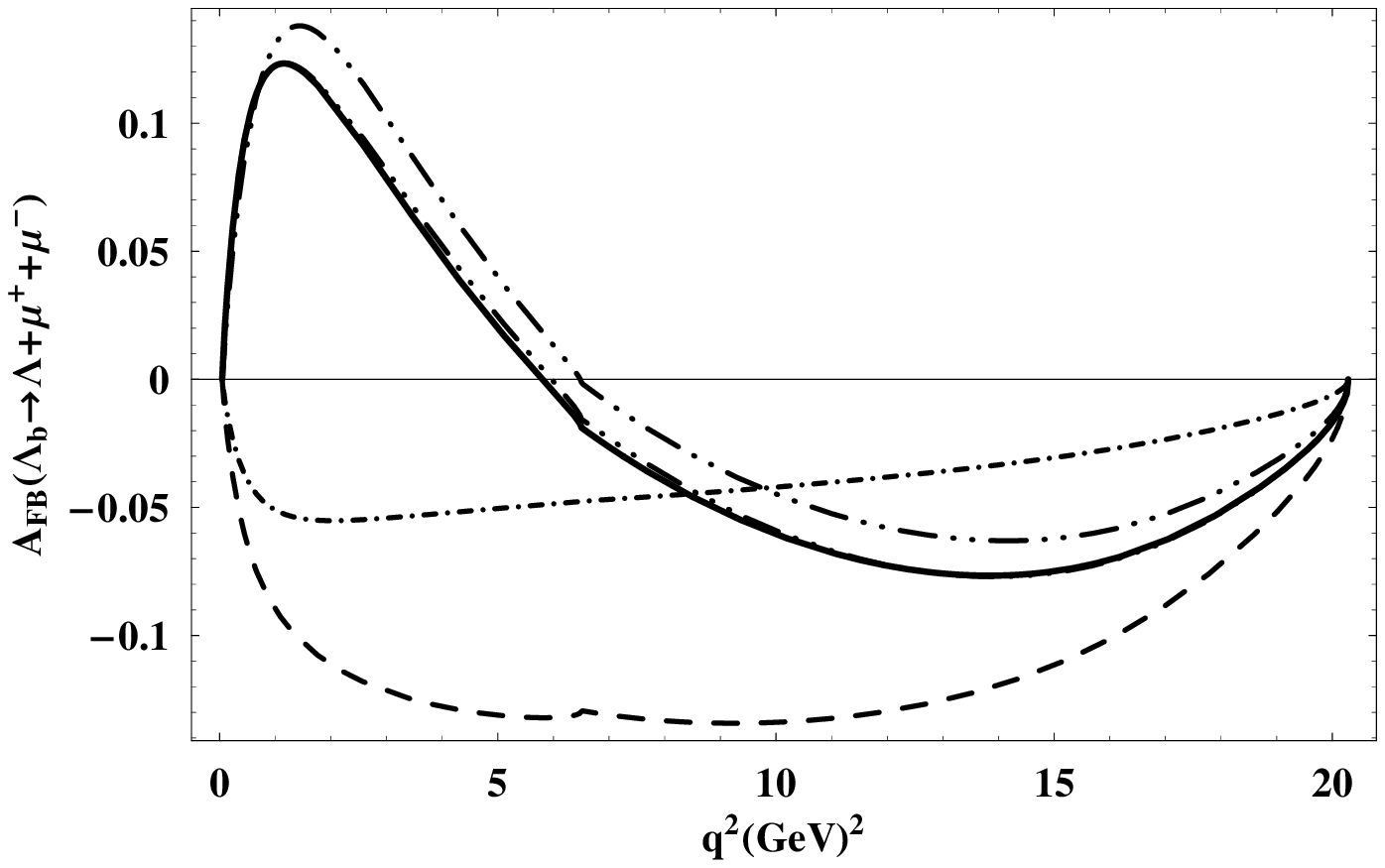} %
\includegraphics[scale=0.6]{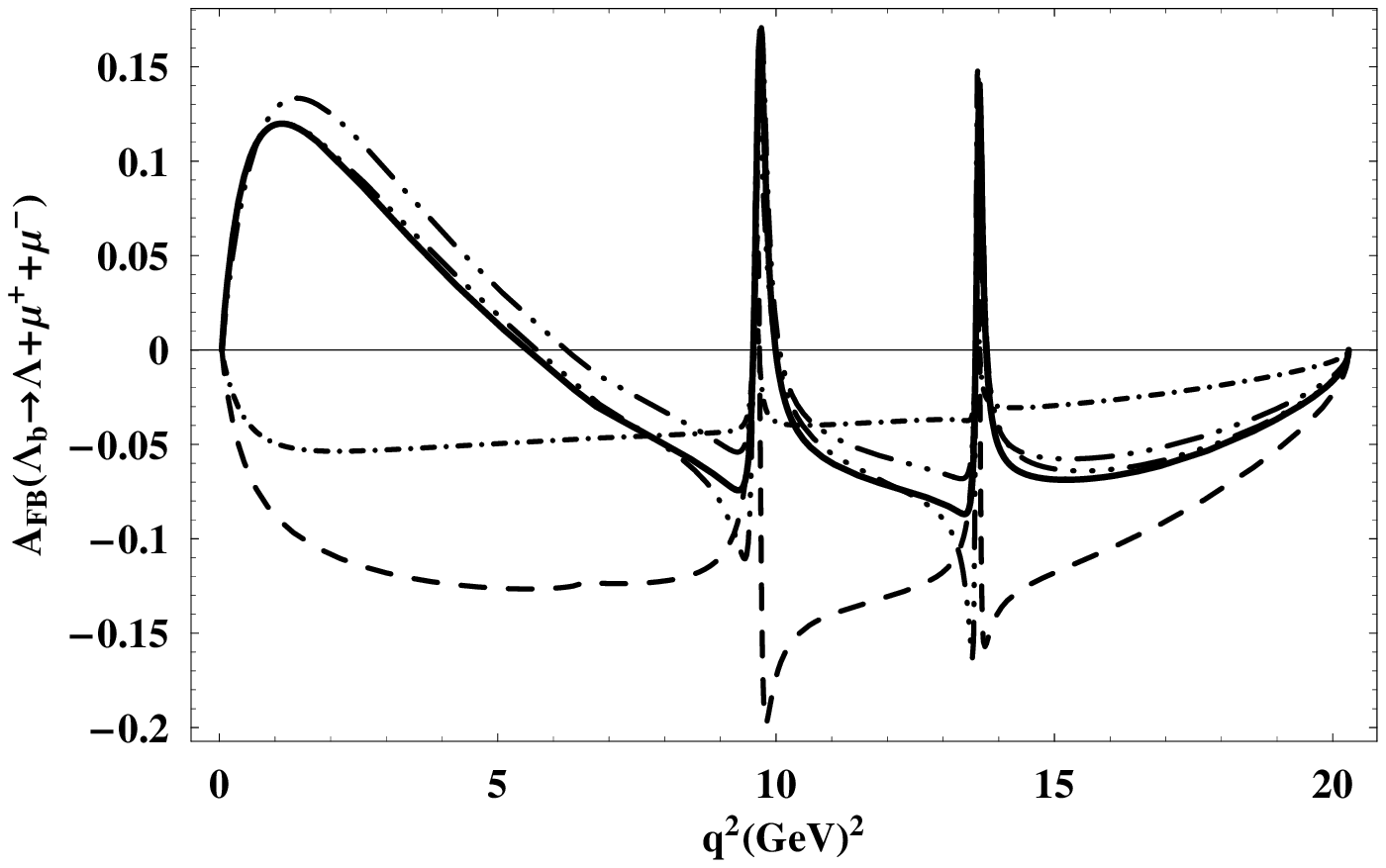} &  &  \\
\includegraphics[scale=0.6]{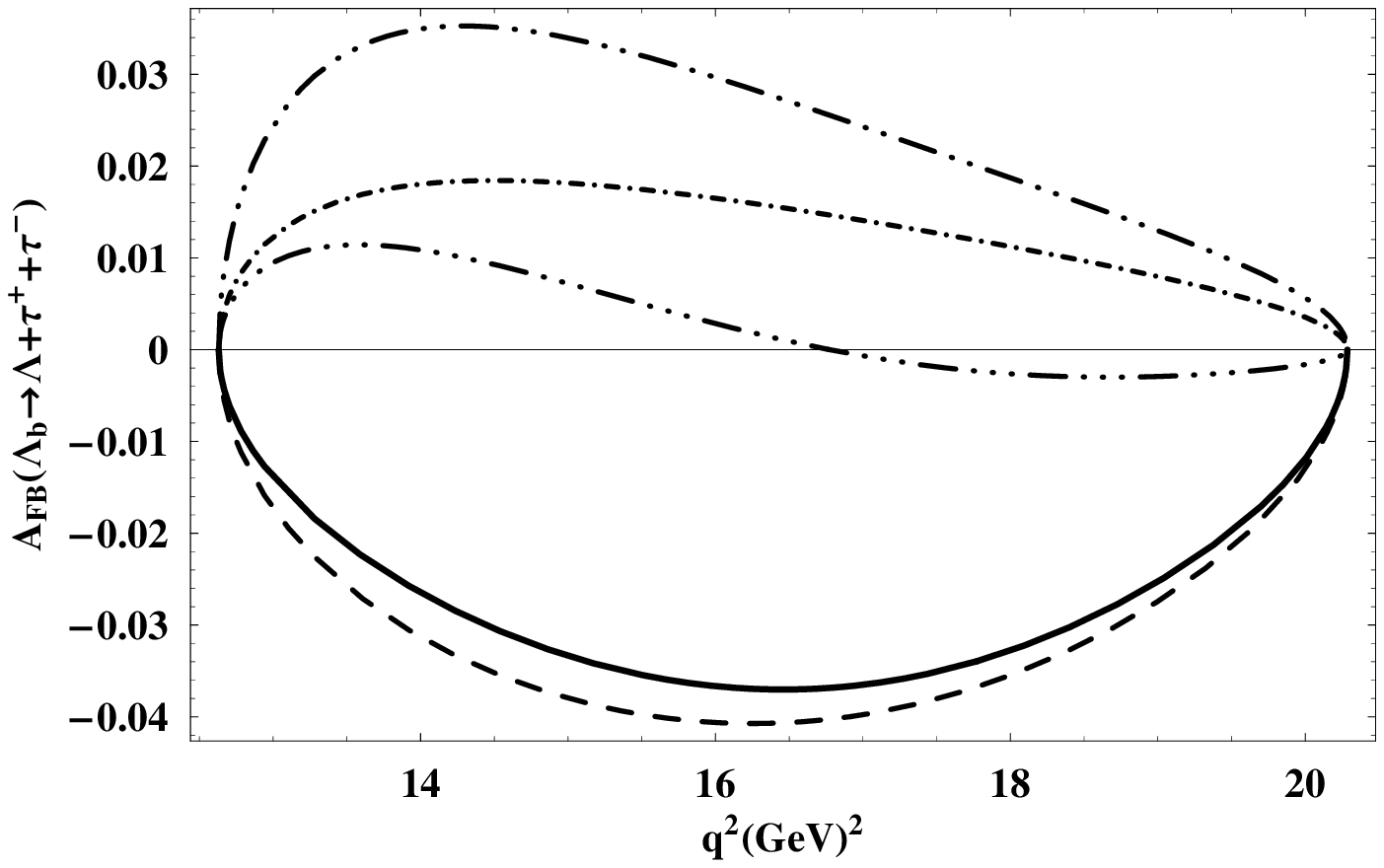} \includegraphics[scale=0.6]{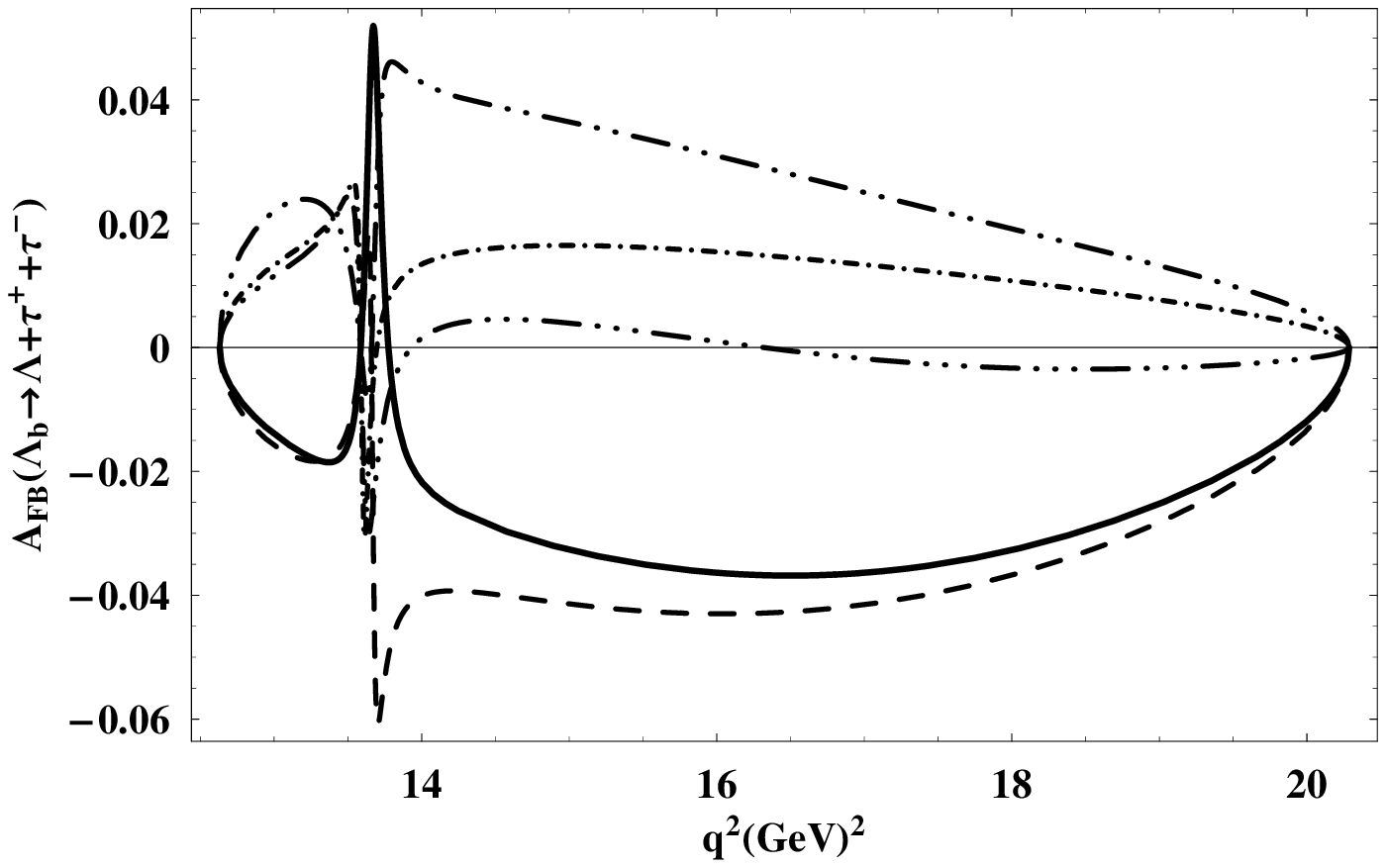}
\put (-350,220){(a)} \put (-100,220){(b)} \put (-350,30){(c)}
\put(-100,30){(d)} \vspace{-1cm} &  &
\end{tabular}%
\end{center}
\caption{Forward-backward asymmetry for the $\Lambda_b \to \Lambda l^+l^-$ ($%
l=\protect\mu, \protect\tau$) decays as functions of $q^2$ without
long-distance contributions (a, c) and with long-distance contributions (b,
d). The line conventions are same as given in the legend of Fig. 1. }
\label{forward backward asymmetry}
\end{figure}

\begin{figure}[tbp]
\begin{center}
\begin{tabular}{ccc}
\vspace{-2cm} \includegraphics[scale=0.6]{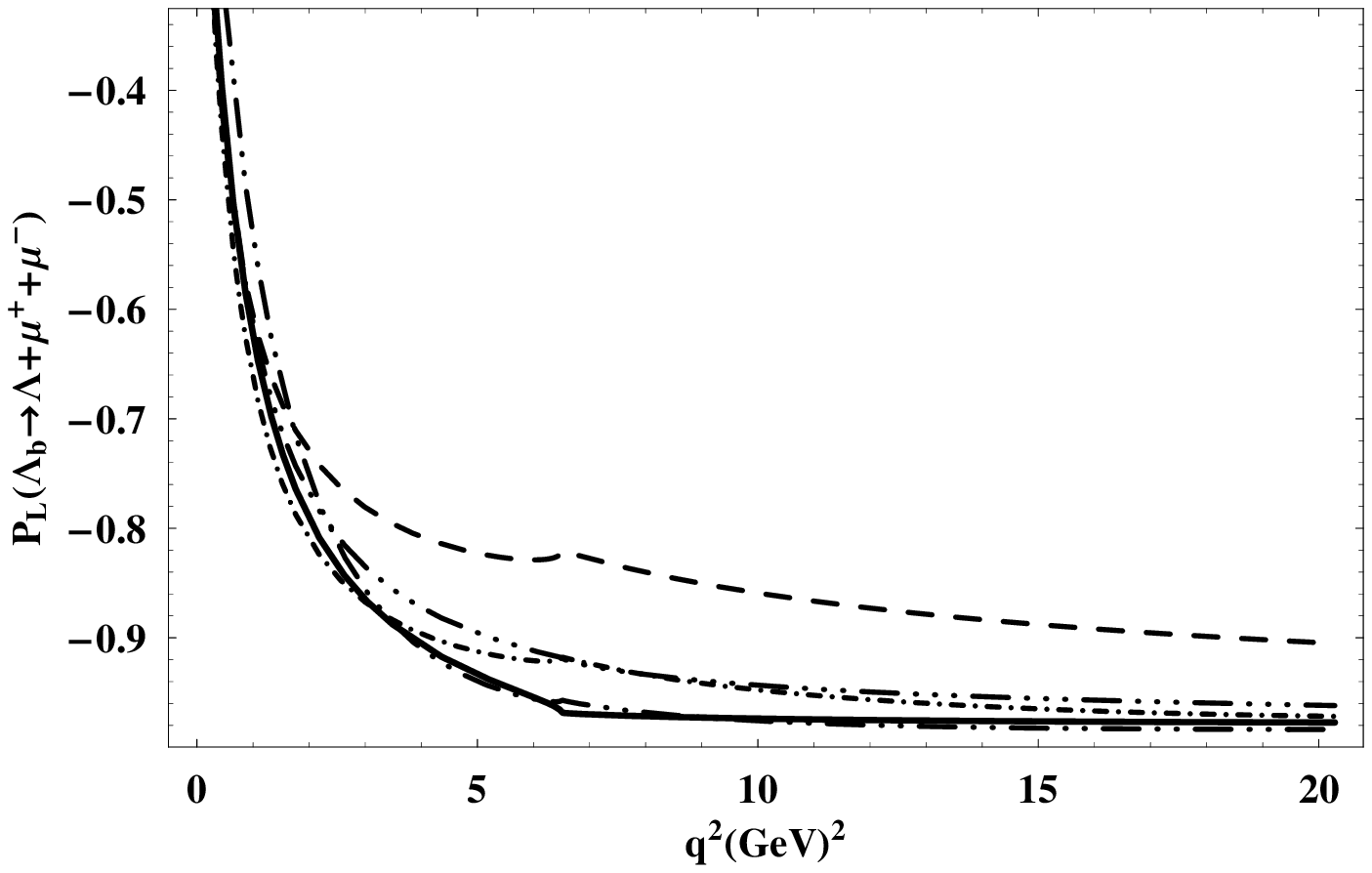} %
\includegraphics[scale=0.6]{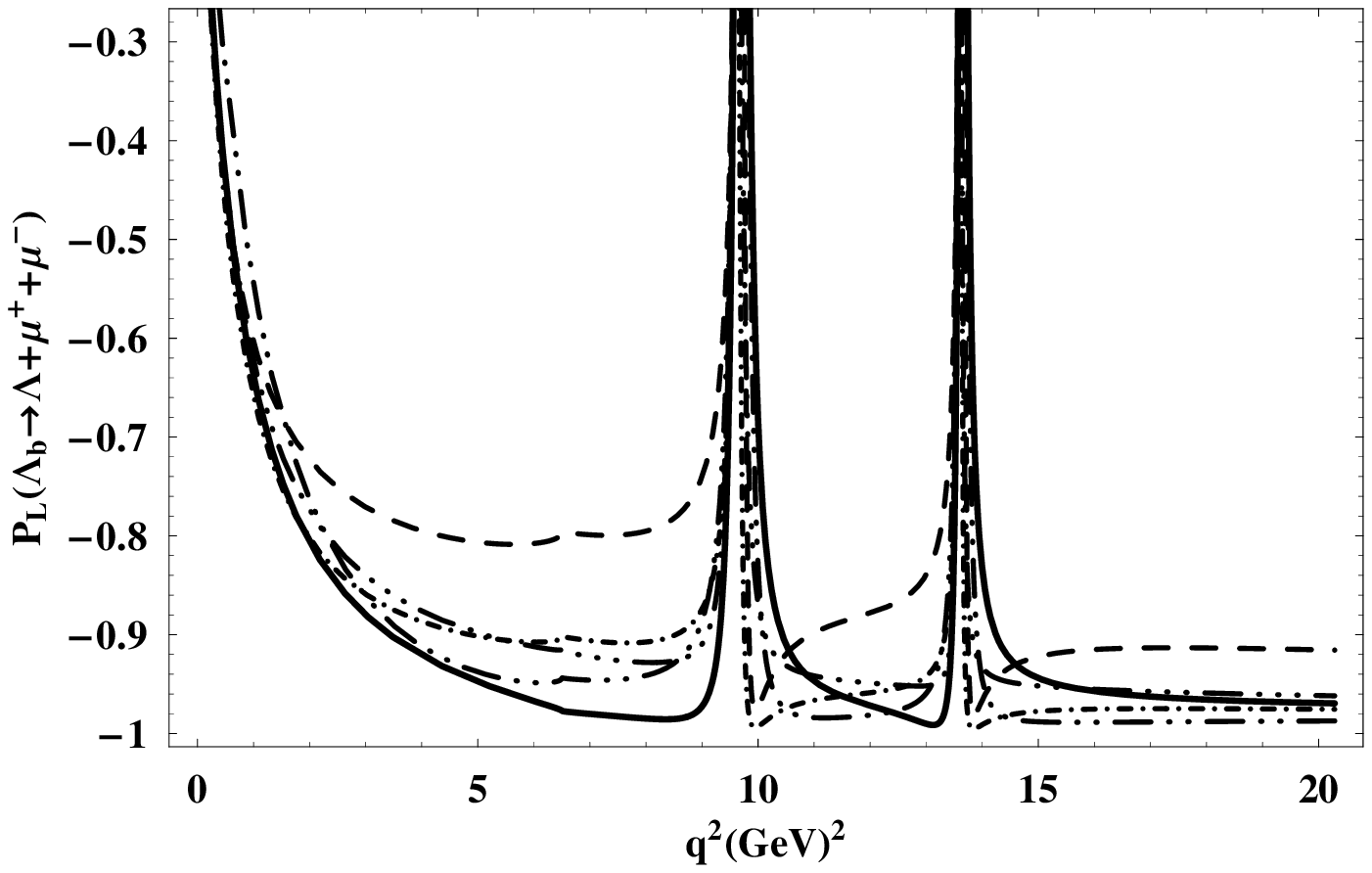} &  &  \\
\includegraphics[scale=0.6]{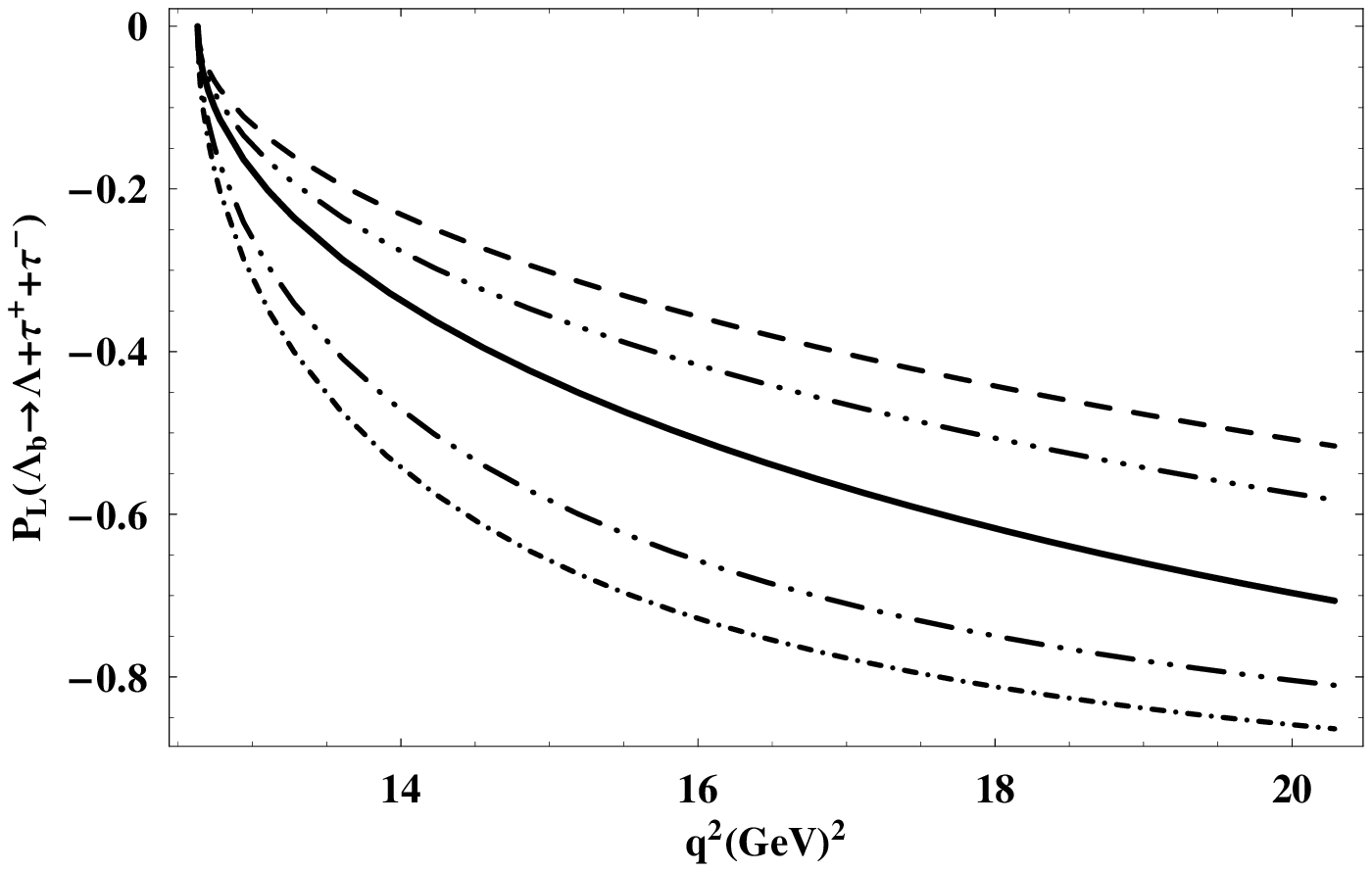} \includegraphics[scale=0.6]{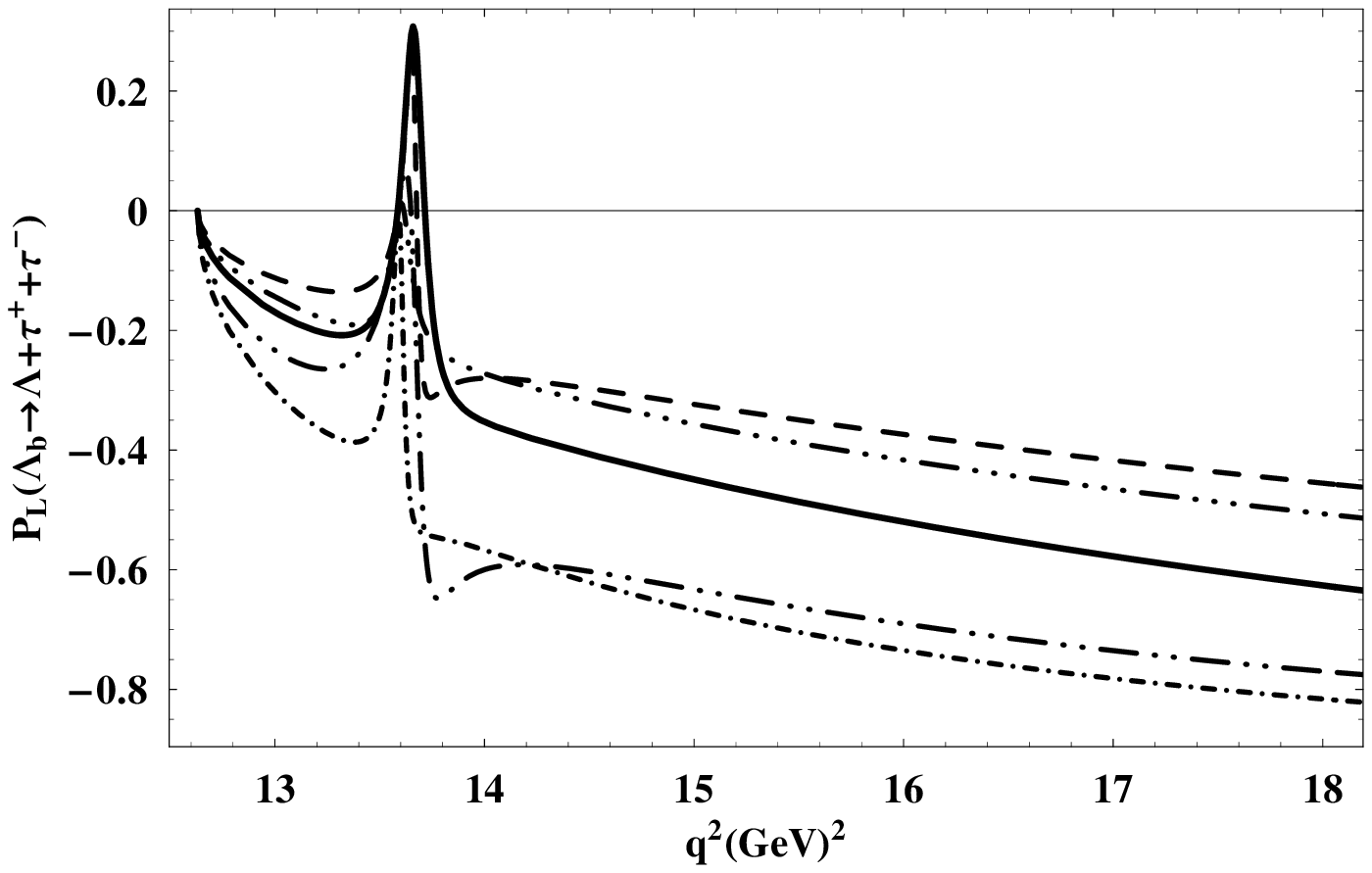}
\put (-350,220){(a)} \put (-100,220){(b)} \put (-350,30){(c)}
\put(-100,30){(d)} \vspace{-1cm} &  &
\end{tabular}%
\end{center}
\caption{Longitudinal lepton polarization asymmetries for the
$\Lambda_b \to \Lambda l^+l^-$ ($l=\protect\mu, \protect\tau$)
decays as functions of $q^2$ without long-distance contributions (a,
c) and with long-distance contributions (b, d). The line conventions
are same as given in the legend of Fig. 1 } \label{Longitudinal
lepton polarization asymmetry}
\end{figure}

In Fig. 2, the FBAs for $\Lambda_{b} \rightarrow \Lambda l^{+}l^{-}$
are presented. Fig. 2(a,b) describe the FBAs for $\Lambda
_{b}\rightarrow \Lambda \mu ^{+}\mu ^{-}$ with and without
long-distance contributions, from which one can easily distinguish
different SUSY models. It is known that in the SM the zero position
of FBAs is due to the opposite sign of $C_{7}^{eff}$ and
$C_{9}^{eff}$. In SUSY I and SUSY II models, the sign of
$C_{7}^{eff}$ and $C_{9}^{eff}$ are the same and hence the zero
point of the FBAs disappears. Whereas, in SUSY III model due to the
opposite sign of $C_{7}^{eff}$ and $C_{9}^{eff}$ forward-backward
asymmetry passes from the zero but this zero position shifts to the
right from that of the SM value due to the contribution from the
NHBs. Similar behavior is expected in SUSY SO(10) GUT model but in
this case the shifting is very mild as the contribution from the
NHBs is very small. For $\Lambda _{b}\rightarrow \Lambda \tau
^{+}\tau ^{-}$ the FBAs with and without long-distance contributions
are represented in Fig. 2(c,d). Again, one can easily distinguish
between the contributions from different SUSY models. Here, the most
interesting point is that the FBAs pass through zero point in SUSY
SO(10) GUT model. This is due to the same sign of the
$C_{Q_{1}}^{\prime }$ and $C_{Q_{2}}^{\prime }$ which
suppress the large contribution coming from the $C_{7}^{eff}$ and $%
C_{9}^{eff}$ in this model. Though SUSY effects are more
distinguishable in FBAs in this case, however, it is too difficult
to measure it experimentally due to its small value.
\begin{figure}[tbp]
\begin{center}
\begin{tabular}{ccc}
\vspace{-2cm} \hspace{-1cm}
\includegraphics[scale=0.50]{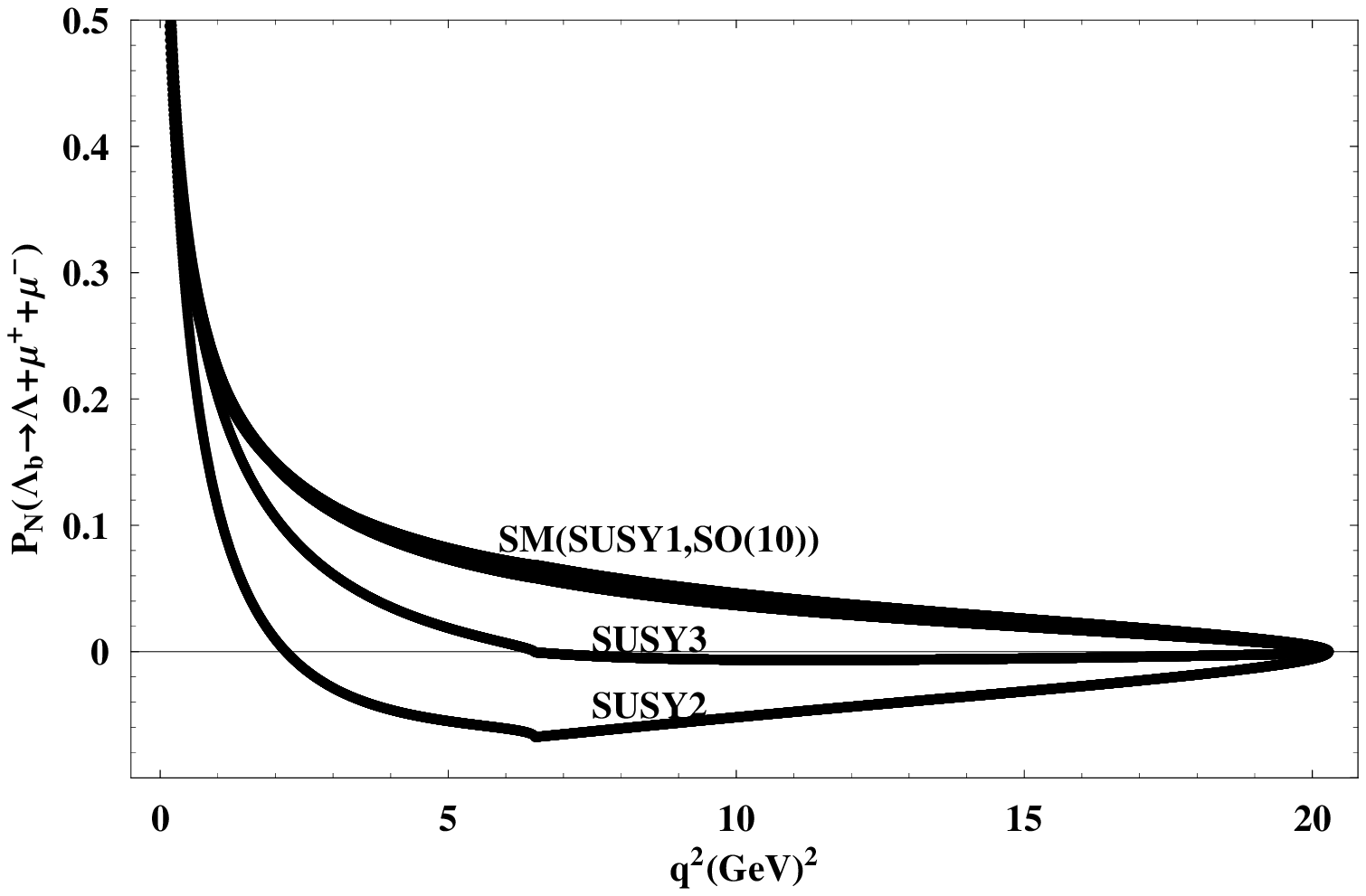}
\hspace{-2cm}
\includegraphics[scale=0.50]{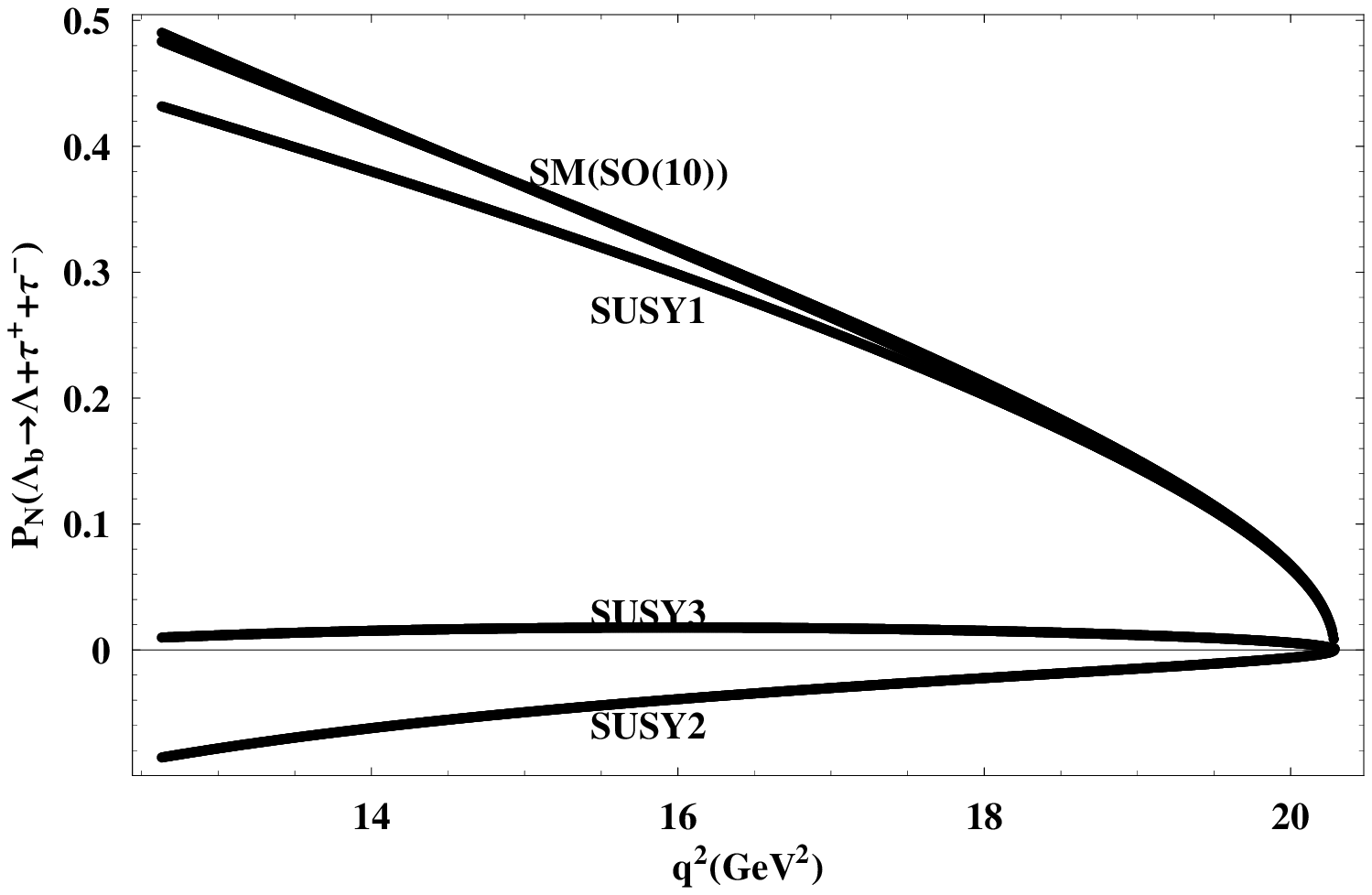}
\put (-400,20){(a)} \put(-150,20){(b)} \vspace{1 cm}
\end{tabular}%
\end{center}
\caption{Normal lepton polarization asymmetries for the $\Lambda_b
\to \Lambda l^+l^-$ ($l=\protect\mu, \protect\tau$) decays as
functions of $q^2$. } \label{Normal lepton polarization asymmetry}
\end{figure}

Figs. (3-5) describe the lepton polarization asymmetries for
$\Lambda _{b}\rightarrow \Lambda l^{+}l^{-}$. Before, we try to
explain the behavior of different polarization asymmetries with the
help the formulas (\ref{expression-LP}, \ref{expression-PN},
\ref{expression-TP}) given above. Eq. (\ref{expression-LP}) shows
the dependence of the longitudinal lepton polarization on different
Wilson coefficients, from which one can expect that the value of
lepton polarization asymmetries in the SUSY I model should be
greatly modified from that of the SM due to the change of sign for
the term proportional to $C_{7}^{*eff}C_{10}$. Due to this change in
sign, the large positive contribution comes and the magnitude of the
longitudinal polarization asymmetry decreases from that of the SM
value. However, this value is expected to increase in SUSY II model
because of the
NHBs contribution which lies in the first and second term of Eq. (\ref%
{expression-LP}). In SUSY III model, this asymmetry lies close to
that of the SM value due to the same sign of $C_{7}^{*eff}C_{10}$
and  small contribution from the NHBs. As we have considered all the
primed Wilson coefficients to be zero therefore the effect of SUSY
SO(10) on longitudinal lepton polarization asymmetry will be
explained by plotting it with the square of the momentum transfer.

Now, Fig. 3(a,b) shows the dependence of longitudinal polarization asymmetry
for the $\Lambda_b \to \Lambda \mu^+\mu^-$ on the square of momentum
transfer. The value in SUSY I model is significantly different from that of
the SM, however, this value is close to that in the SM for SUSY II and SUSY
III models. Furthermore, the absolute value of longitudinal polarization
asymmetry in the SUSY SO(10) is small compared to the SM model due to the
complex part of the Wilson coefficients and also due to small contributions
of the NHBs in this model.

\begin{figure}[tbp]
\begin{center}
\begin{tabular}{ccc}
\hspace{-2cm} \includegraphics[scale=0.50]{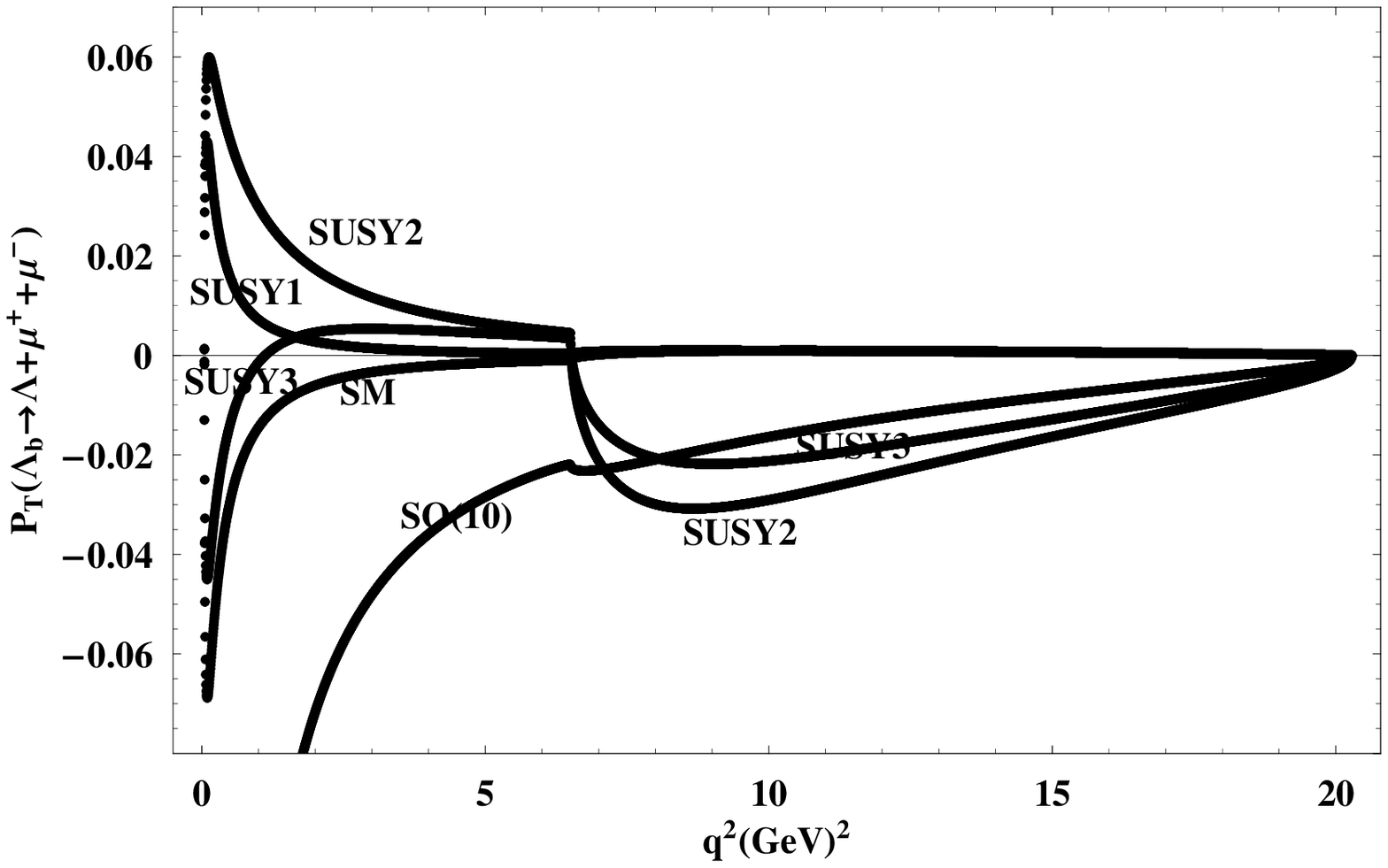} \vspace{-5cm}
\\ \hspace{-2cm}
\includegraphics[scale=0.60]{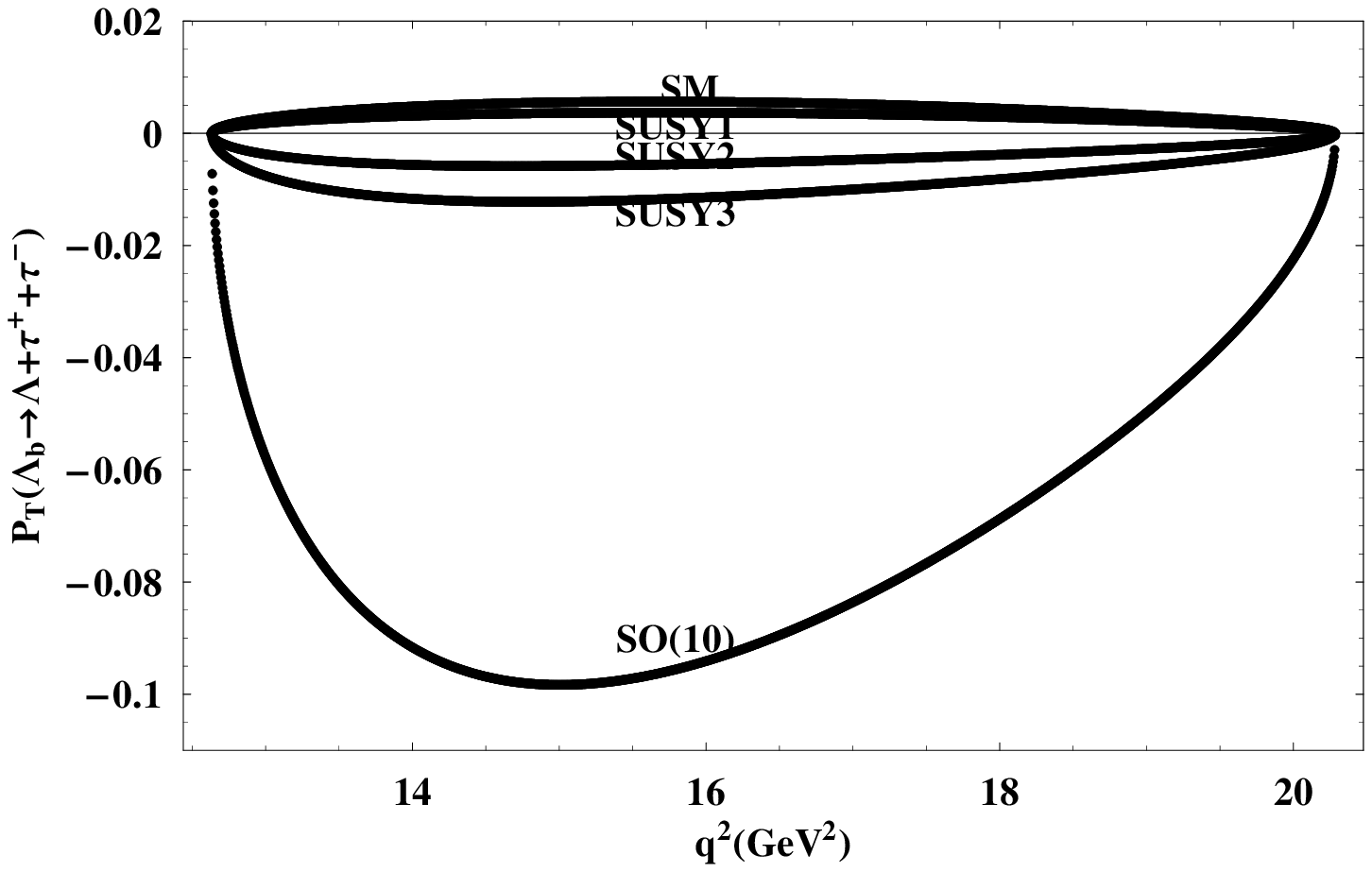} \put (-160,400){(a)} \put
(-160,160){(b)}  \vspace{-5cm}
\end{tabular}%
\end{center}
\caption{Transverse lepton polarization asymmetries for the
$\Lambda_b \to \Lambda l^+l^-$ ($l=\protect\mu, \protect\tau$)
decays as functions of $q^2$. } \label{Transverse lepton
polarization asymmetry}
\end{figure}

Fig. 3(c,d) are for the longitudinal lepton polarization asymmetries of $%
\Lambda _{b}\rightarrow \Lambda \tau^{+}\tau^{-}$ with and without
long-distance contributions, where the different SUSY models are
easily distinguishable. Contrary to the muon case, the values of
this asymmetry in SUSY II and SUSY III models are even larger in
magnitude than that obtained in the SM owing to the large
contributions form NHBs which is attributed to the first and second
term of Eq. (\ref{expression-LP}). Though the large contributions
come from the first term which is proportional to $m_{l}$ in Eq.
(\ref{expression-LP}), but this is overshadowed by the much larger
term proportional to $m_{\Lambda _{b}}(C_{Q_{1}}^{\ast
}C_{Q_{2}}+C_{Q_{2}}^{\ast }C_{Q_{1}})$.

The dependence of lepton normal polarization asymmetries for
$\Lambda _{b}\rightarrow \Lambda l^{+}l^{-}$ on the momentum
transfer square are presented in Fig. 4. In terms of Eq.
(\ref{expression-PN}), one can observe that this asymmetry is
sensitive to the contribution of NHBs in SUSY\ II and SUSY\ III\
models, while it is insensitive to the contributions from SUSY I and
SUSY\ SO(10) model. It can be seen that $P_{N}$ changes its sign in
the case of large contributions from NHBs as indicated in Fig. 4 and
this is also clear from the first three terms of Eq.
(\ref{expression-PN}). As expected, the contribution of NHBs from
the $\tau ^{+}\tau ^{-}$ channel is much more significant than that
from the $\mu ^{+}\mu ^{-}$ channel. Now, the normal polarization is
proportional to the $\lambda $ which approaches to zero at large
momentum transfer region and hence the normal polarization is
suppressed by $\lambda $ in this region.

\begin{figure}[tbp]
\begin{center}
\begin{tabular}{ccc}
\vspace{-1cm} \includegraphics[scale=0.56]{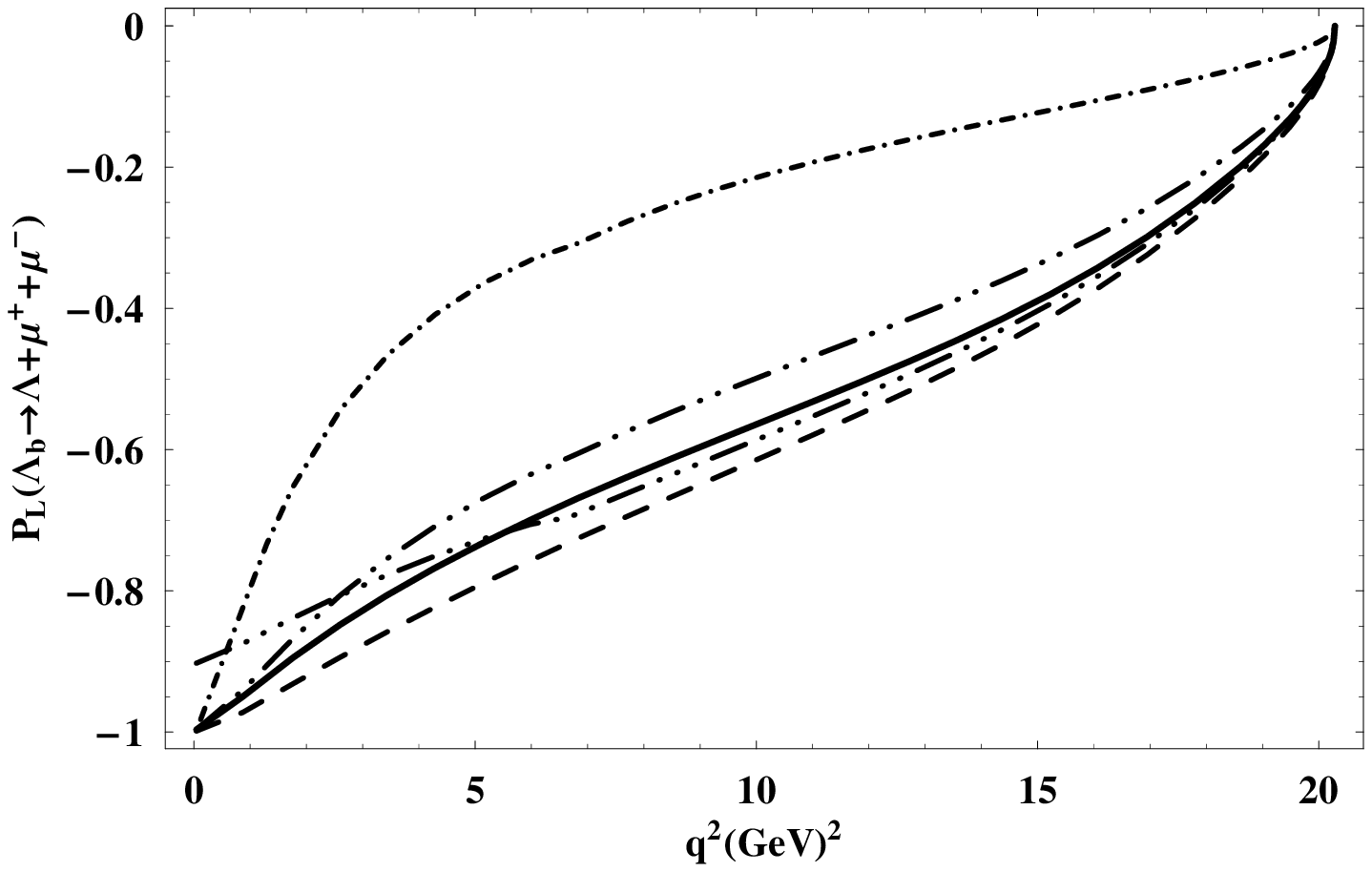} %
\includegraphics[scale=0.54]{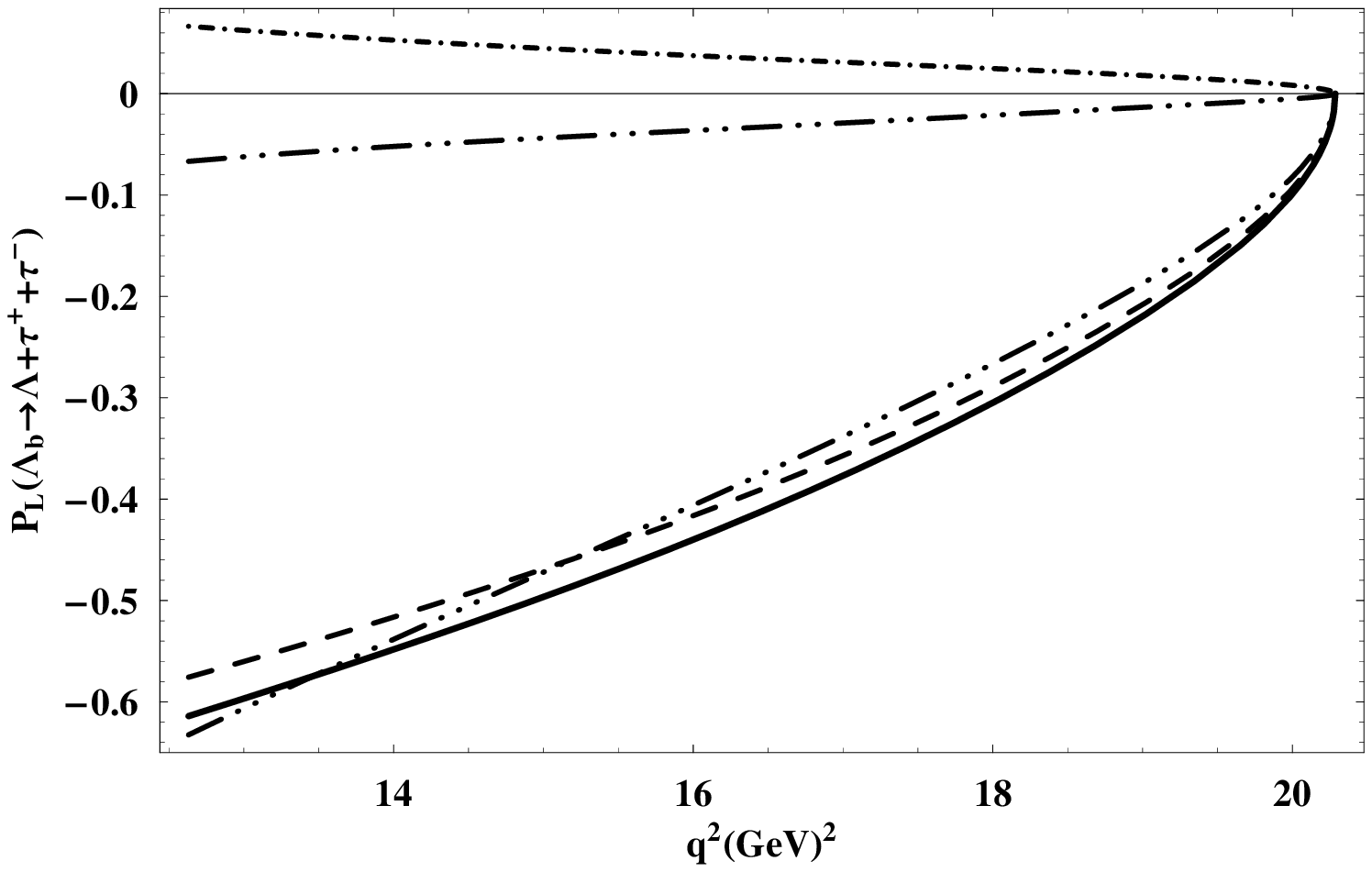} \put (-350, 20){(a)} \put
(-100,20){(b)} &  &
\end{tabular}%
\end{center}
\caption{Longitudinal $\Lambda $ polarization asymmetries for the
$\Lambda_b \to \Lambda l^+l^-$ ($l=\protect\mu, \protect\tau$)
decays as functions of $q^2$. } \label{Normal lepton polarization
asymmetry}
\end{figure}

Fig. 5 shows the dependence of transverse polarization asymmetries for $%
\Lambda _{b}\rightarrow \Lambda l^{+}l^{-}$ on the square of
momentum transfer. From Eq. (\ref{expression-TP}) we can see that it
is proportional to the imaginary part of the Wilson coefficient
which are negligibly small in SM as well as in SUSY I, SUSY II and
SUSY III models. However, complex flavor non-diagonal down-type
squark mass matrix elements of 2nd and 3rd generations are of order
one at GUT scale in SUSY SO(10) model, which induce complex
couplings and Wilson coefficients. As a result, non zero transverse
polarization asymmetries for $\Lambda _{b}\rightarrow \Lambda
l^{+}l^{-}$ exist in this model. Even though in this case, the
asymmetry effects are quite small in $\tau ^{+}\tau ^{-}$ channel,
but the value of transverse polarization asymmetry can reach to
$-0.1$ when the momentum transfer is around $15 \mathrm{GeV^2}$.
Experimentally, to measure $\left\langle P_{T}\right\rangle $ of a
particular decay branching ratio $\mathcal{B}$ at the $n\sigma $
level, the required number of events are
$N=n^{2}/(\mathcal{B}\left\langle P_{T}\right\rangle ^{2})$ and if
$\left\langle P_{T}\right\rangle \sim 0.1$, then the required number
of events are almost $10^{8}$ for $\Lambda _{b}$ decays. Since at
LHC and BTeV machines, the expected number of $b\bar{b}$ production
events is around $10^{12}$ per year, so the measurement of
transverse polarization asymmetries in the $\Lambda _{b}\rightarrow
\Lambda l^{+}l^{-}$ decays could discriminate the SUSY SO(10) model
from the SM and other SUSY models.

Fig. 6 shows the dependence of longitudinal polarization of $\Lambda $
baryon on the square of momentum transfer. One can see that the effects of
NHBs are quite distinguishable in SUSY\ II and SUSY III models both for the $%
\mu ^{+}\mu ^{-}$ and $\tau ^{+}\tau ^{-}$ channels; but the values for SUSY
I and SUSY II are almost close to that of the SM. As observed from Eq. (\ref%
{polarizationlambda-LP}), the effects of NHBs are proportional to
the mass of leptons, therefore the large deviations from the SM are
expected for the tauon as presented in Fig. 6. In the SUSY II model,
the value of the longitudinal polarization even changes its sign for
$\tau ^{+}\tau ^{-}$ channel.

\begin{figure}[tbp]
\begin{center}
\begin{tabular}{ccc}
\hspace{-1.5cm} \includegraphics[scale=0.50]{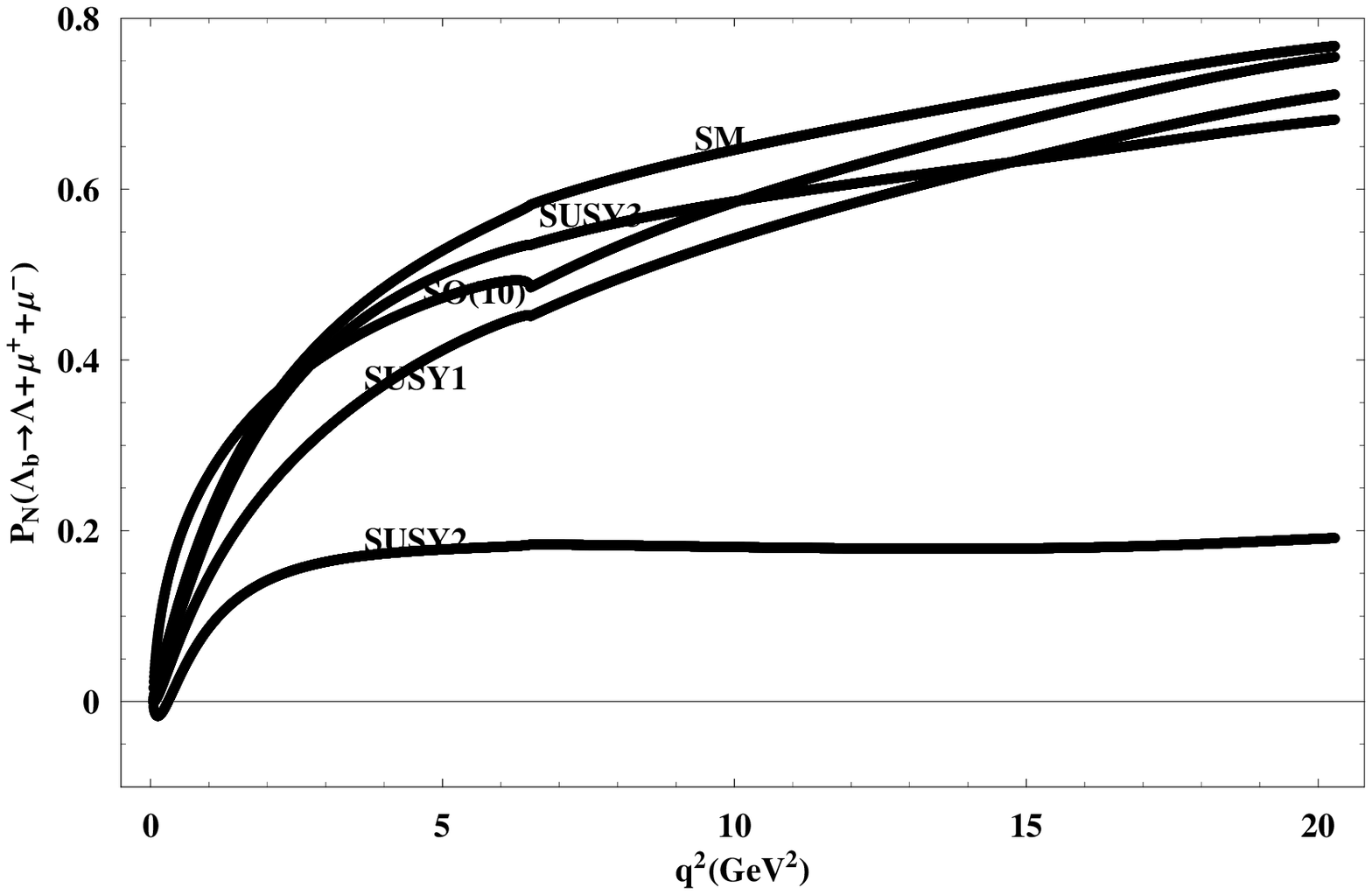} %
\hspace{-1.5cm}\includegraphics[scale=0.52]{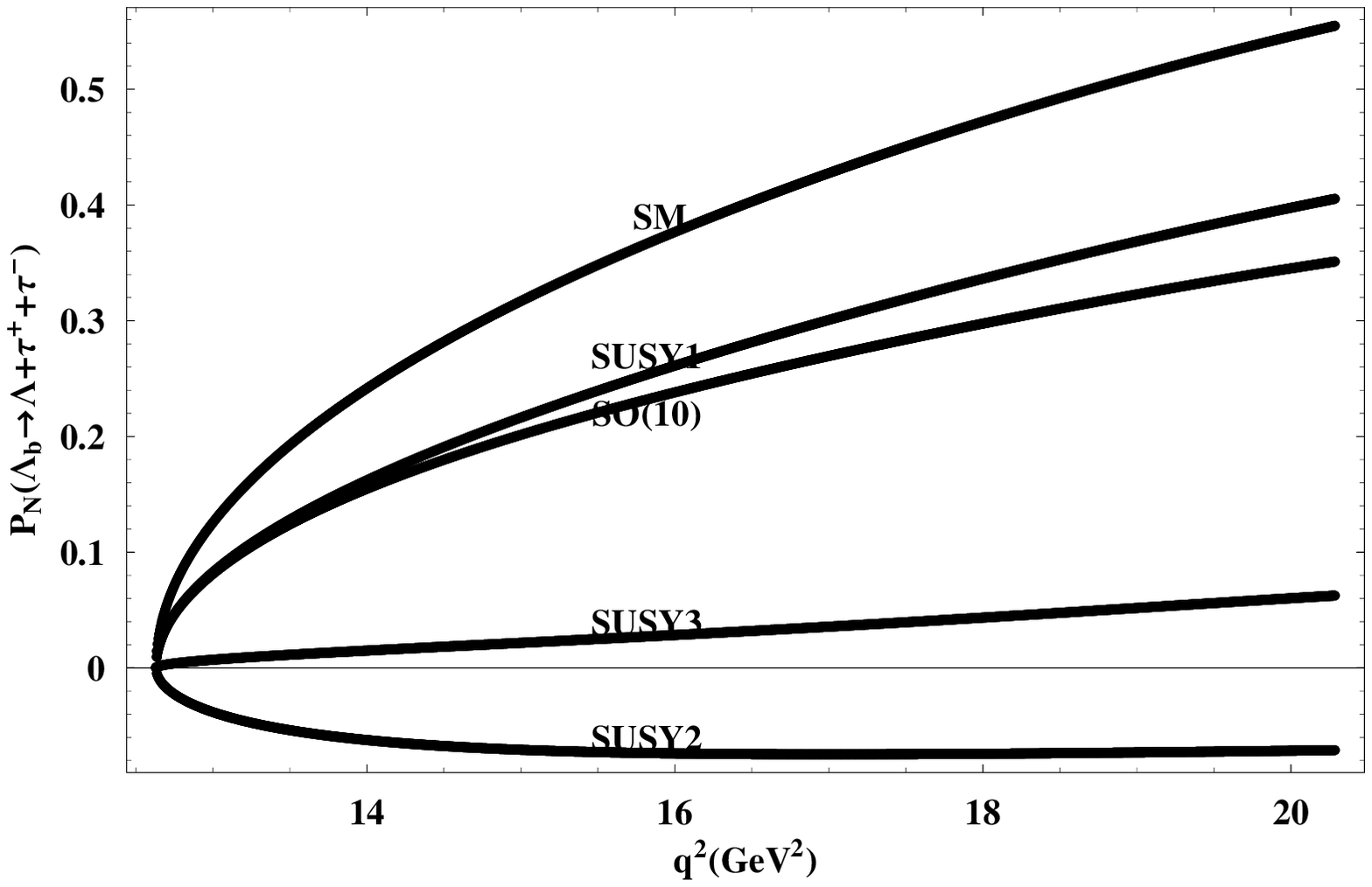} \put
(-400,100){(a)} \put (-120,100){(b)} &  & \vspace{-4cm}
\end{tabular}%
\end{center}
\caption{Normal $\Lambda $ polarization asymmetries for the
$\Lambda_b \to \Lambda l^+l^-$ ($l=\protect\mu, \protect\tau$)
decays as functions of $q^2$. } \label{Normal lepton polarization
asymmetry}
\end{figure}

With the help of Eq. (\ref{lambdapolarization-PN}), one can see that the
normal polarization asymmetry of $\Lambda $ baryon is sensitive to the $%
C_{Q_{1}}$ and the sign of the $C_{7}^{eff}$. It is shown that the sign of $%
C_{7}^{eff}$ is negative in SUSY I and II models. In particular, this
asymmetry in the SUSY II model differs from that in the SM remarkably due to
the large value of $C_{Q_{1}}$. Moreover, the contributions from SUSY III
and SUSY SO(10) models are also quite distinguishable from the SM as shown
in Fig. 7. Therefore, the measurements of normal polarization asymmetries for $%
\Lambda _{b}\rightarrow \Lambda l^{+}l^{-}$ in future experiments will help
to distinguish different scenarios beyond the SM .

\begin{figure}[tbp]
\vspace{-2cm}
\begin{center}
\begin{tabular}{ccc}
\hspace{-2.5cm}\includegraphics[scale=0.60]{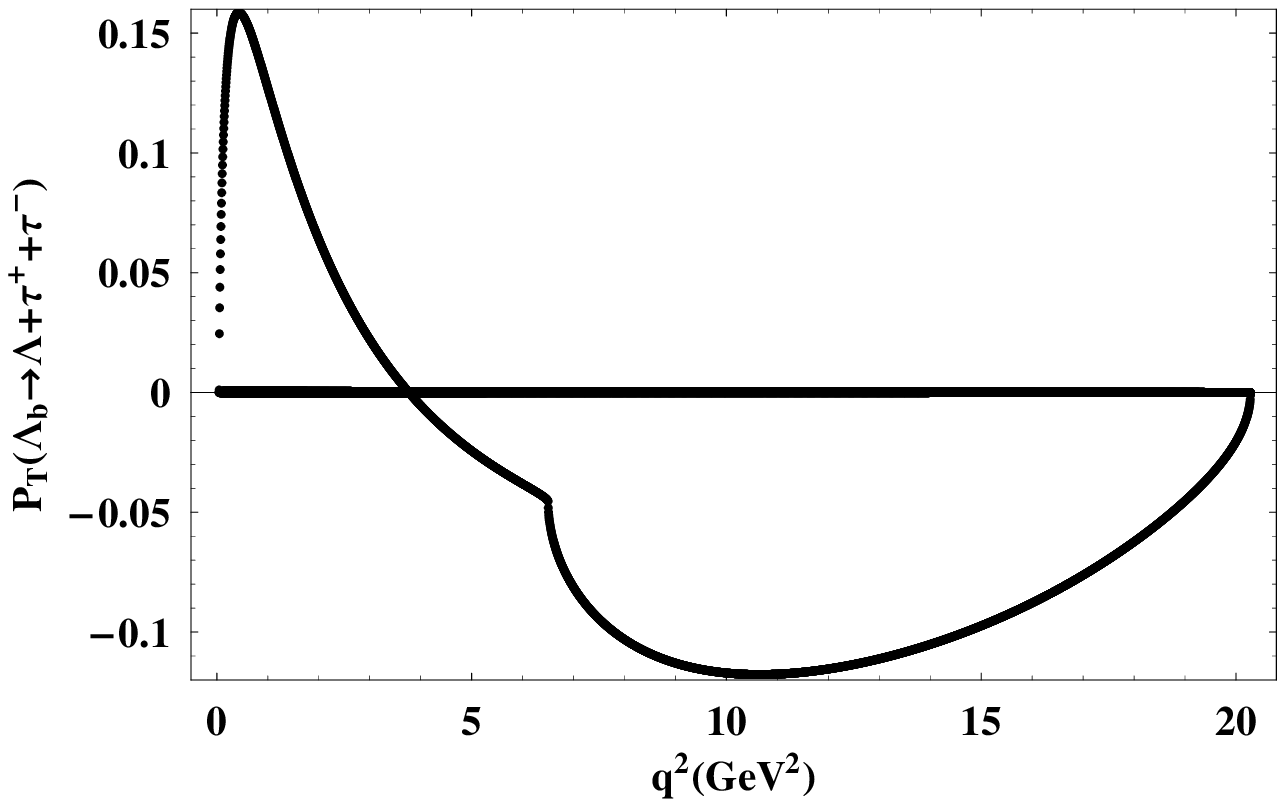} %
\hspace{-3cm}
\includegraphics[scale=0.58]{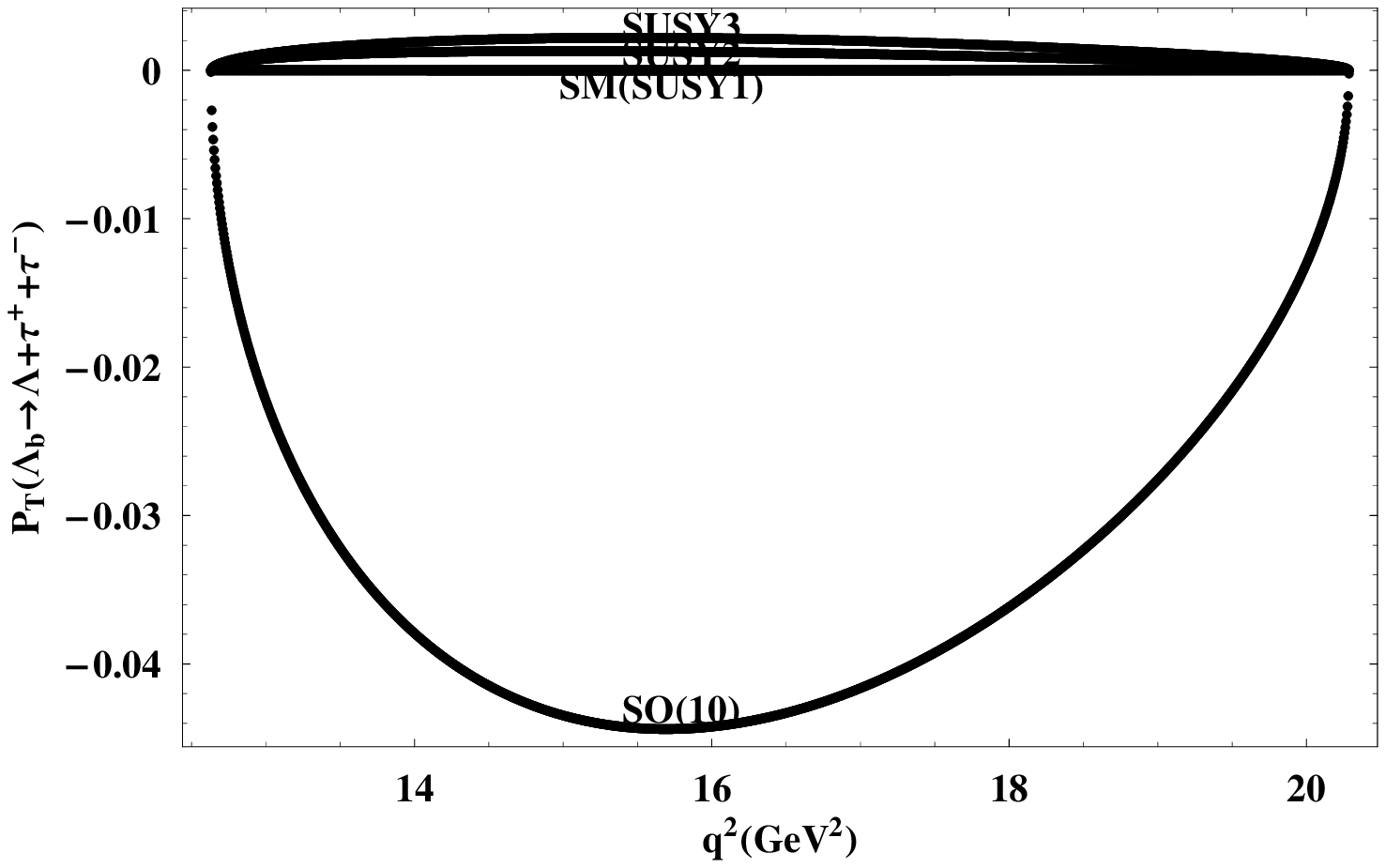} \put (-420,135){(a)} \put
(-150,135){(b)} &  & \vspace{-5cm}
\end{tabular}%
\end{center}
\caption{Transverse $\Lambda $ polarization asymmetries for the
$\Lambda_b \to \Lambda l^+l^-$ ($l=\protect\mu, \protect\tau$)
decays as functions of $q^2$. }
\end{figure}

Similar to the transverse polarization asymmetry of lepton, the
transverse polarization asymmetry of $\Lambda $ baryon is also
proportional to the imaginary part of $C_{Q_{1}}C_{7}^{*eff}$ and
$C_{Q_{1}}C_{9}^{*eff}$ (c.f. Eq. (\ref{Lambdapolarization-PT})). It
is known that these imaginary parts are quite small in SM, SUSY I,
SUSY\ II and SUSY\ III\ model and hence the values of the transverse
polarization asymmetries of $\Lambda $ baryon are almost zero in
these models. However, the imaginary part of Wilson coefficient in
SUSY SO(10) model is large and hence its effects are quite different
from the other models discussed above as collected in Fig. 8. For
the muon case, the transverse polarization asymmetry can reach the
number $-0.1$ in SUSY SO(10) model; whereas the value is too small
to measure experimentally for the tauon case.

\section{Conclusion}

We have carried out the study of invariant mass spectrum, FBAs,
polarization asymmetries of lepton and $\Lambda$ baryon in $%
\Lambda _{b}\rightarrow \Lambda l^{+}l^{-}$ ($l=\mu ,\tau $) decays in SUSY
theories including SUSY SO(10) GUT model. Particularly, we analyze the
effects of NHBs to this process and our main outcomes can be summarized as
follows:

\begin{itemize}
\item The differential decay rates deviate sizably from that of the SM
especially in the large momentum transfer region. These effects are
significant in SUSY II model where the value of the Wilson
coefficients corresponding to the NHBs is large. However, the SUSY
SO(10) effects in differential decay rate of $\Lambda
_{b}\rightarrow \Lambda l^{+}l^{-}$ are negligibly small .

\item The SUSY effects show up for the FBA of the $\Lambda _{b}\rightarrow
\Lambda l^{+}l^{-}$ ($l=\mu ,\tau $) decays and the deviations from
the SM are very large especially in SUSY I and SUSY II model where
the FBAs did not
pass from zero. The reason is the same sign of $C_{7}^{eff}$ and $%
C_{9}^{eff} $ in these two models, but in SUSY III scenario it
passes from the zero-point different from that of the SM. The
effects of SUSY SO(10) are quite distinguishable when the final
state leptons are the tauon pair but these are too small to be
measured experimentally.

\item The longitudinal, normal and transverse polarizations of leptons are
calculated in different SUSY models. It is found that the SUSY effects are
very promising which could be measured at future experiments and shed light
on the new physics beyond the SM.

\item Following the same line, the longitudinal, normal and transverse
polarizations of $\Lambda $ baryon in $\Lambda _{b}\rightarrow
\Lambda l^{+}l^{-}$ ($l=\mu ,\tau $) decays are calculated at
length. The different SUSY effects are clearly distinguishable from
each other and also from that of the SM. The transverse polarization
asymmetries are proportional to the imaginary part of the Wilson
coefficients. Hence it is almost zero in SM as well as in MSSM
model, however, the Wilson coefficients in SUSY SO(10) GUT model
have large imaginary part and hence the value of the transverse
polarization is expected to be non-zero. The maximum value of
transverse polarization of $\Lambda _{b}\rightarrow \Lambda
\mu^{+}\mu^{-}$ decay reaches to $-0.1$ for the square of momentum
transfer around $10 \mathrm{GeV^2} $ and hence can be measurable at
future experiments like LHC and BTeV machines where a large number
of $\ b \bar {b} $ pairs are expected to be produced.
\end{itemize}

In short, the experimental investigation of observables, like decay
rates, FBAs, lepton polarization asymmetries and the polarization
asymmetries of $ \Lambda $ baryon in $\Lambda _{b}\rightarrow
\Lambda l^{+} l^{-}$ ($l=\mu, \tau$) decay will be used to search
for the SUSY effects, in particular the NHBs effect, encoded in the
MSSM as well as SUSY SO(10) models.

\section*{Acknowledgements}

This work is partly supported by National Science Foundation of
China under Grant No.10735080 and 10625525. The authors would like
to thank Cheng Li, Yue-Long Shen and Qi-Shu Yan for helpful
discussions.

\end{document}